\documentclass[iop]{emulateapj}

\usepackage{amsmath, amsthm, amssymb}
\usepackage{color}
\usepackage{nicefrac}         

\newcommand {\nbodypp}     {\textsc{\mbox{nbody6\raise.2ex\hbox{\tiny{++}}}}}
\newcommand {\starlab}     {\texttt{starlab}}

\newcommand {\Msun}        {\mbox{M$_{\odot}$}}

\newcommand {\kms}         {\mbox{km/s}}
\newcommand {\pc}          {\mbox{pc}}

\newcommand {\pcdens}      {\mbox{pc$^{-3}$}}

\newcommand {\mdnorm}      {\mbox{$\hat{m}_{\mathrm{d}}$}}
\newcommand {\Dmdnorm}     {\mbox{$\Delta{\hat{m}_{\mathrm{d}}}$}}
\newcommand {\Dmd}         {\mbox{$\delta{m_{\mathrm{d}}}$}}
\newcommand {\tailstars}   {\mbox{${\cal{T}}^\bullet$}}
\newcommand {\tailstarsA}  {\mbox{${\cal{T}}^\bullet_\mathrm{A}$}}
\newcommand {\tailstarsB}  {\mbox{${\cal{T}}^\bullet_\mathrm{B}$}}
\newcommand {\tailstarsAB} {\mbox{${\cal{T}}^\bullet_\mathrm{AB}$}}
\newcommand {\tailstarsC}  {\mbox{${\cal{T}}^\bullet_\mathrm{C}$}}
\newcommand {\tailstarsD}  {\mbox{${\cal{T}}^\bullet_\mathrm{D}$}}

\begin{document}

\title{The Evolution of Protoplanetary Disks in the Arches Cluster}

\shorttitle{Discs in the Arches Cluster}

\shortauthors{Olczak et al.}

\author{C. Olczak\altaffilmark{1,2,3}}
\affil{Astronomisches Rechen-Institut (ARI), Zentrum f{\"u}r Astronomie Universit{\"a}t Heidelberg, M{\"o}nchhofstrasse 12-14, 69120 Heidelberg, Germany}
\email{olczak@ari.uni-heidelberg.de}

\author{T. Kaczmarek}
\affil{Max-Planck-Institut f{\"u}r Radioastronomie, Auf dem H{\"u}gel 7, 53121 Bonn, Germany}

\author{S. Harfst}
\affil{Technische Universit{\"a}t Berlin, Zentrum f{\"u}r Astronomie und Astrophysik, Hardenbergstra{\ss}e 36, 10623 Berlin, Germany}

\author{S. Pfalzner}
\affil{Max-Planck-Institut f{\"u}r Radioastronomie, Auf dem H{\"u}gel 7, 53121 Bonn, Germany}

\and

\author{S. Portegies Zwart}
\affil{Sterrewacht Leiden, Leiden University, Postbus 9513, 2300 RA Leiden, The Netherlands}

\altaffiltext{1}{Max-Planck-Institut f{\"u}r Astronomie (MPIA), K{\"o}nigstuhl 17, 69117 Heidelberg, Germany}
\altaffiltext{2}{National Astronomical Observatories of China, Chinese Academy of Sciences (NAOC/CAS), 20A Datun Lu, Chaoyang District, Beijing 100012, China}
\altaffiltext{3}{The Kavli Institute for Astronomy and Astrophysics at Peking University (KIAA), Yi He Yuan Lu 5, Hai Dian Qu, Beijing 100871, China}

\begin{abstract}
  Most stars form in a cluster environment. These stars are initially surrounded by discs from which potentially planetary systems form. Of all
  cluster environments starburst clusters are probably the most hostile for planetary systems in our Galaxy. The intense stellar radiation and extreme
  density favour rapid destruction of circumstellar discs via photoevaporation and stellar encounters. Evolving a virialized model of the Arches
  cluster in the Galactic tidal field we investigate the effect of stellar encounters on circumstellar discs in a prototypical starburst
  cluster. Despite its proximity to the deep gravitational potential of the Galactic centre only a moderate fraction of members escapes to form an
  extended pair of tidal tails. Our simulations show that encounters destroy one third of the circumstellar discs in the cluster core within the first
  2.5\,Myr of evolution, preferentially affecting the least and most massive stars. A small fraction of these events causes rapid ejection and the
  formation of a weaker second pair of tidal tails that is overpopulated by disc-poor stars. Two predictions arise from our study: (i)~If not
  destroyed by photoevaporation protoplanetary discs of massive late B- and early O-type stars represent the most likely hosts of planet formation in
  starburst clusters. (ii)~Multi-epoch $K$- and $L$-band photometry of the Arches cluster would provide the kinematically selected membership sample
  required to detect the additional pair of disc-poor tidal tails.
\end{abstract}



%

\section{Introduction}

\label{sec:introduction}

Observations in the past decade have shown that most young stars do not form in isolation but as part of a cluster environment
\citep[e.g.][]{2003ARA&A..41...57L,2003AJ....126.1916P,2009ApJS..181..321E}. The accretion discs of these stars are thus exposed to environmental
effects that could affect their evolution
\citep{1998A&A...340..508R,2000prpl.conf..401H,2001MNRAS.325..449S,2004ApJ...611..360A,2006ApJ...642.1140O,2006ApJ...652L.129P,2006A&A...454..811P,2007A&A...462..193P,2007MNRAS.376.1350C,2010A&A...509A..63O}.
As these discs are the prerequisites for the formation of planetary systems this process might be influenced by the cluster environment. However, so
far theoretical investigations about the effect of irradiation by massive stars or strong gravitational interactions have concentrated on low- and
intermediate mass star clusters like IC348 or the Orion Nebula Cluster (ONC). Much more massive systems like NGC~3603, the Arches cluster or
Westerlund~1 - known as starburst clusters - that are expected to trigger the strongest effects have not been treated. Considering the huge amount of
massive stars and extreme densities, i.e. more than 100 O-stars and a core density $\gtrsim$$10^5\,\Msun\,\pcdens$ in the Arches cluster
\citep{1999ApJ...525..750F}, an extrapolation of the previous investigations towards starburst clusters would suggest that the lifetime of
circumstellar discs could be shortened dramatically and so planet formation in such an environment hindered significantly.

However, recent observations of the Arches cluster in the near-infrared $JKL'$ bands by \citet{2010ApJ...718..810S} have revealed emission from
circumstellar matter around at least a few percent of stars in the mass range $2-20\,\Msun$. This detection of discs in the Arches B-star population
was surprising for two reasons. First, it is expected that the stars' UV radiation causes depletion of their own inner disc. Considering that the
characteristic UV evaporation timescale of a primordial disc around Herbig Be stars is less than 1\,Myr \citep{2009A&A...497..117A}, a disc lifetime
of 2.5\,Myr for B-type stars implies that the self-photoerosion of discs is probably less efficient than various models suggest
\citep[e.g.][]{2000prpl.conf..401H,2008NewAR..52...60A}. Second, extrapolating from the above mentioned numerical studies of less massive clusters, in
a starburst cluster environment disc destruction is expected to be much increased by external processes. The extreme UV radiation from numerous
O-stars and gravitational interactions induced by the high stellar density could boost the removal of external disc material.

In the present investigation we focus on the mechanism of encounter-induced disc-destruction to constrain its contribution to the overall disc life
time. We use simulations of the Arches cluster preformed by the $\starlab$ simulation package
\citep[][]{1996ASPC...90..413M,2001MNRAS.321..199P,2003IAUS..208..331H} and carry out the analysis of the disc-mass loss analogous to our previous
publications referenced above. Hence this study of a massive cluster complements previous investigations of encounter-induced disc evolution in low-
and intermediate-mass clusters
\citep{2001MNRAS.325..449S,2004ApJ...611..360A,2006ApJ...642.1140O,2006ApJ...652L.129P,2006A&A...454..811P,2007A&A...462..193P,2007MNRAS.376.1350C,2010A&A...509A..63O}.

In contrast to the low- and intermediate-mass clusters in the solar neighbourhood the location of the massive Arches cluster near the Galactic centre
exposes it to strong tidal fields. Therefore another difference to previous simulations is that our model includes the contribution of the Galactic
tidal field. Stars escaping from a star cluster in the gravitational field of a galaxy form extended tidal tails known from various observations in
the Milky Way \citep{1995AJ....109.2553G,1997A&AS..121..439K,2000A&A...359..907L,2006ApJ...637L..29B} and numerical models
\citep{1999A&A...352..149C,2002ApJ...565..265P,2005AJ....129.1906C,2007MNRAS.380..749F}. Except for high-speed escapers created in few-body
encounters, stars escape as a result of two-body encounters passing at slow speed close to the saddle points of the effective potential, known as
Lagrange points \citep{2008MNRAS.387.1248K}.

With our study we aim to investigate the evolution of the encounter induced-disc mass loss in a tidally distorted starburst cluster and its tidal
arms. Of particular relevance is the work of \citet{2009MNRAS.392..969J} who use the epicycle theory for a quantitative analysis of the tidal tail
structure of star clusters moving on a circular orbit in the Galactic disc. They find that the radial offset of a star's epicyclic motion relative to
the orbit of the star cluster, $\Delta{R_0}$, is first order in the angular momentum, $\Delta{L}$, while its radial amplitude~$r_\mathrm{m}$ depends
on the energy excess, $\Delta{E}$, which is of second order in~$\Delta{L}$:
\begin{equation}
  \label{eq:just09}
  \begin{aligned}
    \Delta{R_0}  &\propto \Delta{L} \,,\\
    r_\mathrm{m} &\propto \sqrt{\Delta{E}} \,.
  \end{aligned}
\end{equation}
\citet{2009MNRAS.392..969J} conclude that the circumstances are more complicated near the Galactic Centre but to lowest order the theory is still
applicable.

Throughout this work we assume that initially all stars are surrounded by protoplanetary discs. This is justified by observations that reveal disc
fractions of nearly 100\,\% in very young star clusters
\citep[e.g.][]{2000AJ....120.1396H,2000AJ....120.3162L,2001ApJ...553L.153H,2005astro.ph.11083H}. The typical disc diameter is a few hundred AU for
low- and intermediate-mass stars \citep{1996AJ....111.1977M,2008Ap&SS.313..119A}, though discs of more than several thousand AU surrounding massive
stars have been observed \citep[see][and references therein]{2005IAUS..227..135Z}. However, it remains unclear whether a clear correlation of disc
extension and stellar mass does exist \citep[see e.g.][]{2005A&A...441..195V}.

In Section~\ref{sec:observations_onc} we outline the observationally determined basic properties of the Arches cluster that serves as a reference for
our cluster models. The computational method and the properties of the numerical models are described in
Section~\ref{sec:numerical_method}. Afterwards we present results from our numerical simulations in Section~\ref{sec:numerical_results}. The
conclusion and discussion mark the last section of this paper.

%

\section{Properties of the Arches cluster}

\label{sec:observations_onc}

The Arches cluster is one of the most massive and densest young clusters in the Galaxy and harbors about 125 O~stars \citep{1999ApJ...525..750F}. Most
of its stars have formed more or less simultaneously about 2.5\,Myr ago \citep{2004ApJ...611L.105N,2008A&A...478..219M}. The cluster is very compact
with a core radius $R_\mathrm{c} \approx 0.15$\,pc \citep{2009A&A...501..563E}, a half-mass radius $R_\mathrm{hm} \approx 0.4$\,pc
\citep{2002ApJ...581..258F,2005ApJ...628L.113S} and a tidal radius $R_\mathrm{t} \gtrsim 1$\,pc \citep{2000ApJ...545..301K}. Based on these size
scales \citet{2009A&A...501..563E} quote a concentration parameter $c = \log{ ( R_\mathrm{t} / R_\mathrm{c} ) } \approx 0.84$. This is equivalent to a
King parameter $W_0 \approx 4$ \citep[e.g.][]{1987gady.book.....B}. In their detailed analysis \citet{2010MNRAS.409..628H} find that stars with $m >
10\,\Msun$ are more concentrated due to mass segregation and best fit by a King profile with $W_0 = 7$.

The Arches cluster is located at a projected distance to the Galactic Center of only about 30\,pc \citep{1995AJ....109.1676N}. Taking into account its
three-dimensional space motion of $232 \pm 30\,\kms$ \citep{2008ApJ...675.1278S}, numerical simulations suggest an eccentric orbit and a true
present-day distance below 200\,pc (\citealp{2008ApJ...675.1278S}; \citealp[see also][]{2002ApJ...565..265P}). The internal motion of the cluster is
characterized by a one-dimensional velocity dispersion of $5.4 \pm 0.4\,\kms$ \citep{2012ApJ...751..132C}. Note that these authors have refined the
bulk proper motion of the Arches cluster that implies a slightly smaller space motion of $196 \pm 17\,\kms$.

It is still under debate whether there is an overabundance of massive stars, i.e. an unusually flat (``top-heavy'') mass function, or if the mass
function is truncated at masses well above the hydrogen burning limit \citep[e.g.][]{2005ApJ...628L.113S}. However, in their most recent work
\citet{2009A&A...501..563E} do not find any significant difference to the less massive star-forming regions in the solar neighbourhood. They derive a
mass of about $3\times10^4$\,$\Msun$ within the estimated half-mass radius \citep[assuming the IMF of][]{2001MNRAS.322..231K} and a core density of
$2\times10^5$\,$\Msun\pcdens$. However, these estimates rely on counts of stellar masses above 10\,$\Msun$ only. The same authors find evidence for
mass segregation in the cluster \citep[e.g.][]{2005ApJ...628L.113S,2009A&A...501..563E}. Despite its young age this feature could be driven
dynamically \citep{1969ApJ...158L.139S,1982ApJ...253..512F} due to the Arches cluster's short half-mass relaxation time of only about 10\,Myr
\citep{2006ApJ...653L.113K,2007MNRAS.378L..29P,2009A&A...501..563E}, though there is no consensus on the significance of the effect
\citep[e.g.][]{2007MNRAS.381L..40D}.

Recently, \citet{2010ApJ...718..810S} determined proper motions of hundreds of stars and simultaneously obtained deep $L'$-band photometry of
OB(A)-type stars for a precise and reliable estimate of the disc fraction. They detect 21 sources in the mass range $\sim$ 2-20\,$\Msun$ with emission
in excess of the expected photospheric flux, indicating the presence of circumstellar dust, and find evidence for three optically thick discs from CO
band head emission. This finding implies a total disc fraction of $6\,\% \pm 2\,\%$ within the observed mass range. The disc fraction increases with
cluster radius from $2.7\,\% \pm 1.8\,\%$ in the core ($r < 0.16\,\pc$) to $5.4\,\% \pm 2.6\,\%$ at intermediate radii ($0.16\,\pc < r < 0.3\,\pc$)
and $9.7\,\% \pm 3.7\,\%$ outside ($r > 0.3\,\pc$).

%

\section{Computational method}

\label{sec:numerical_method}

\subsection{Dynamical cluster model}

\label{sec:numerical_method:cluster}

The dynamical model of the Arches cluster used here is based on the work of \citet{2010MNRAS.409..628H} who compared in detail the results of
numerical simulations with observational data. They were able to constrain the initial conditions and found a best-fitting model as a function of five
initial parameters: (i) mass function slope~$\alpha$, (ii) lower mass limit~$m_\mathrm{low}$, (iii) number of massive stars~$N_\mathrm{MS}$ (with mass
$m > 20\,\Msun$), (iv) virial radius~$R_\mathrm{vir}$, and (v) King parameter~$W_0$. We summarize the main properties of this model.

We assume the cluster is initially (at $t=0$\,Myr) gas-free and in virial equilibrium (i.e. the ratio of kinetic to potential energy is
$Q_\mathrm{vir}$\,=\,0.5) and that its mass is distributed according to a King profile \citep{1966AJ.....71...64K}. A single-aged stellar population
is used and no initial mass segregation is taken into account. Primordial binaries are not included but we discuss their potential dynamical effects
in Section~\ref{sec:conclusions}. A grid of numerical models with varying initial parameters $\alpha$, $m_\mathrm{low}$, $N_\mathrm{MS}$,
$R_\mathrm{vir}$, and $W_0$ was evolved until the present-day age of the Arches cluster, $t_\mathrm{AC} = 2.5$\,Myr, to find the best match with
observational data.

According to the best-fitting models of \citet{2010MNRAS.409..628H} the initial slope~$\alpha$ above $0.5\,\Msun$ is consistent with the IMF of
\citet{2001MNRAS.322..231K}. The models also favour a narrow range for the initial number of massive stars $N_\mathrm{MS} \approx 150-200$. However,
the initial lower mass limit~$m_\mathrm{low}$ is not well constrained by their models and thus only allows to estimate a lower limit for the initial
total mass $M \gtrsim 4 \cdot 10^4\,\Msun$. The initial size of the cluster with $R_\mathrm{vir} \approx 0.7-0.8$\,pc is again well constrained. The
corresponding core radius is about $0.4$\,pc initially, and shrinks, due the dynamical evolution of the cluster, down to the observed $0.2$\,pc. The
best-fitting models have modest initial concentrations with $W_0 = 3-5$. In summary, the favoured fiducial model of the Arches cluster, which we will
use here, is parametrized by $N_\mathrm{MS} = 150$, $R_\mathrm{vir} = 0.7$, and $W_0 = 3$. The total number of particles is $N = 74153$.

Stellar evolution and the orbit of the Arches cluster in the Galactic center potential have been neglected in the simulations of
\citet{2010MNRAS.409..628H}. Here we extend the fiducial model to lower stellar masses ($m_\mathrm{low} = 0.1\,\Msun$) and include these two effects
to generate two models: an ``isolated'' model~I with stellar evolution only and an ``orbital'' model~O with both stellar evolution and a realistic
orbital motion of the cluster in the Galactic center potential.

The simulations were carried out with the direct $N$-body integrator {\tt kira} from the $\starlab$ package
\citep[][]{1996ASPC...90..413M,2001MNRAS.321..199P,2003IAUS..208..331H} that includes modules for stellar evolution and an external gravitational
potential. The stellar evolution module accounts for mass loss by stellar winds and binary evolution
\citep{1989ApJ...347..998E,1996A&A...309..179P,1997MNRAS.291..732T,1998A&A...329..551L}. The external potential that was used in our simulations is
that of a power-law mass function with $M_\mathrm{gal}(r) = 4.25\cdot10^6 (r[\pc])^{1.2}\,\Msun$ \citep{2002ApJ...565..265P} which is based on
observations of the $K$-band luminosity function by \citet{1999A&A...348..457M}. Computations were accelerated using GPUs with the help of the Sapporo
library \citep{2009NewA...14..630G}.

The cluster orbit in the potential is given by the six phase-space coordinates at a given time. Five present-day ($t_\mathrm{AC} = 2.5$\,Myr)
coordinates of the Arches cluster are well known from observations: its line-of-sight velocity \citep[$95\,\kms$:][]{2002ApJ...581..258F}, proper
motion \citep[$190\,\kms$ anti-parallel to the Galactic plane:][]{2008ApJ...675.1278S}, and projected position \citep[$30\,\pc$ from the Galactic
center:][]{1995AJ....109.1676N}. We define a coordinate system in which the $x$-axis is along the Galactic plane, the $y$-axis along the
line-of-sight, and the $z$-axis towards the Galactic north pole, to obtain the following present-day position and velocity vectors of the cluster:
\begin{equation}
  \begin{aligned}
    \boldsymbol{r}_\mathrm{cluster} &= (-24, d_\mathrm{los}, 10)\,\pc \,,\\
    \boldsymbol{\upsilon}_\mathrm{cluster} &= (-190, 95, 0)\,\kms \,,
  \end{aligned}
\end{equation}
where we adopt $d_\mathrm{los} = -100\,\pc$ as a probable value for the line-of-sight distance of the Arches cluster to the Galactic centre
\citep{2008ApJ...675.1278S}. Throughout the paper we use $R_\mathrm{GC} = 8$\,kpc for the Sun's distance to the Galactic center
\citep{2008ApJ...689.1044G,2009ApJ...707L.114G}.

\begin{figure*}
  \centering
  \includegraphics[width=0.32\linewidth]{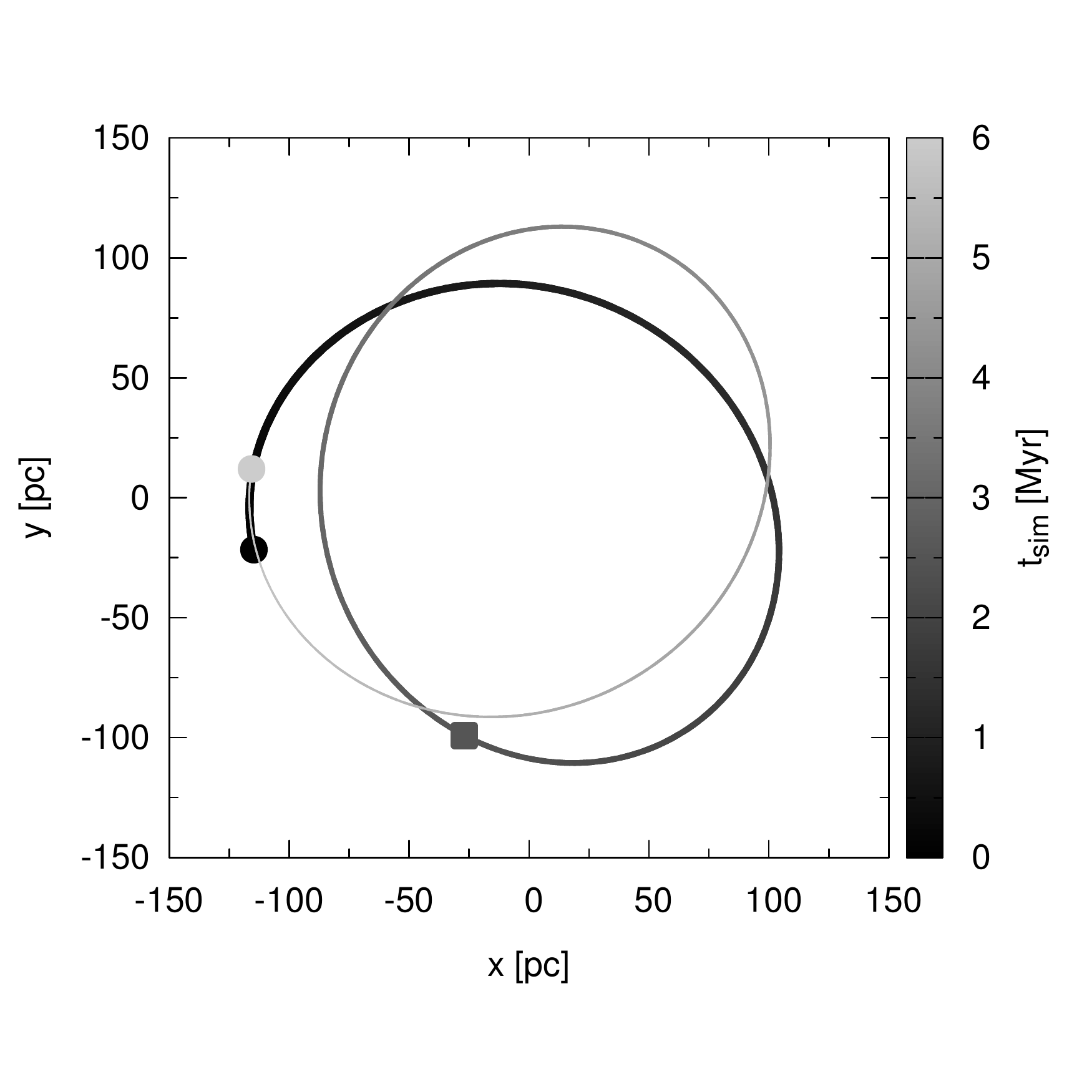}
  \includegraphics[width=0.32\linewidth]{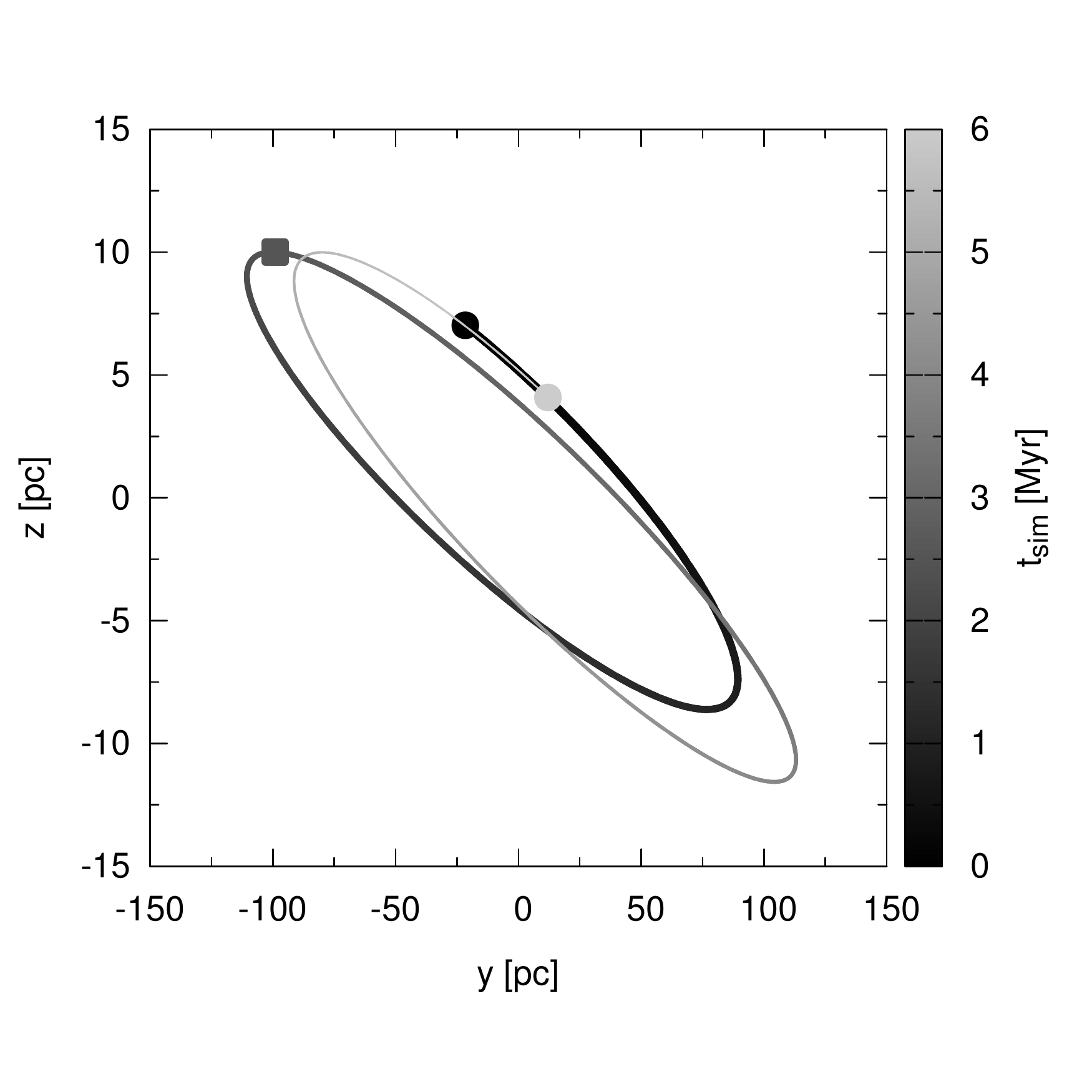}
  \includegraphics[width=0.32\linewidth]{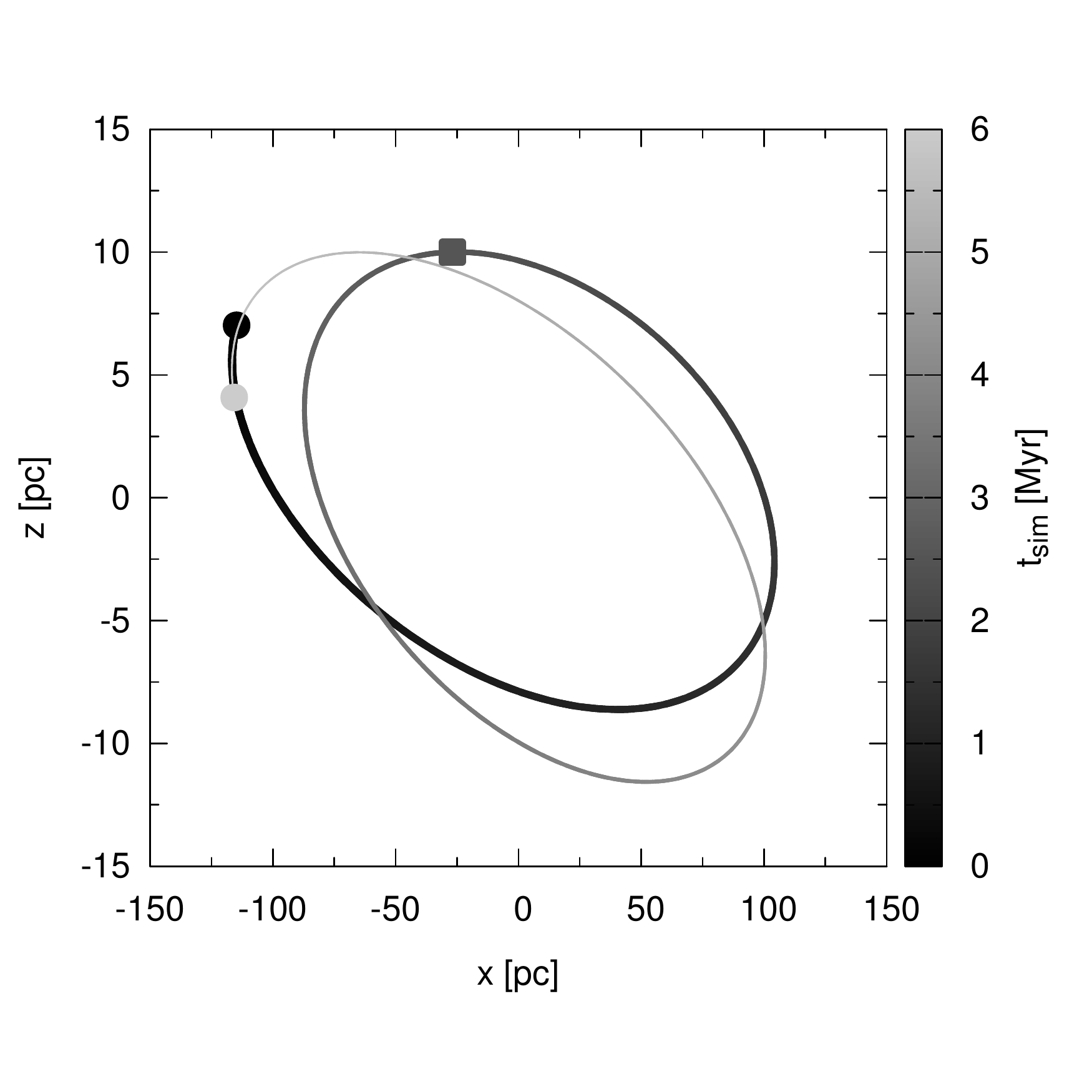}
  \caption{Three planar projections of the orbital motion of the cluster's centre-of-mass for model~O. The origin marks the Galactic centre. The grey
    scale and the line thickness decode the time evolution over 6\,Myr; the black/grey filled circles represent the initial/final position,
    respectively. The position at the present-day age of the Arches cluster ($t_\mathrm{AC} = 2.5$\,Myr) is indicated by the dark grey filled
    rectangle.}
  \label{fig:orbit}
\end{figure*}

The present-day phase space coordinates were used to numerically integerate the orbit backwards in time to find the initial position and velocity of
the Arches cluster. From there, the full cluster was then integrated with {\tt kira} for 6\,Myr including the effects from stellar evolution. Its
orbital motion is shown in Fig.~\ref{fig:orbit}.

We point out that model~I is the best-fitting initial model of the present-day Arches cluster from our set of \emph{isolated} models. However,
computational limitations did not allow to carry out a second parameter study of \emph{orbital} models. Hence model~O is just a modification of
model~I that is evolved on a realistic orbit around the Galactic Centre in the Galactic tidal field. As we will show later in
Section~\ref{sec:numerical_results:cluster_dynamics} this modification has a strong effect on the internal dynamics of model~O such that it becomes a
less good model of the internal dynamics of the Arches cluster compared to model~I.

\subsection{Star-disc encounters}

\label{sec:numerical_method:encounters}

The key aspect of this work is the determination of the encounter-induced disc-mass loss in a starburst cluster environment. For this purpose two
different types of numerical simulations have been combined. First, as part of previous work a parametrized fit of the relative disc-mass loss has
been derived from a tree-code based parameter study of isolated star-disc encounters
\citep{2005A&A...437..967P,2006A&A...454..811P,2010A&A...509A..63O}. Using this fit formula, in a second step the individual relative disc-mass loss
of star cluster members was determined from direct $N$-body simulations of star cluster dynamics, in which the encounter history of all stars was
tracked. We outline the procedure in more detail below.

Stellar encounters in dense clusters can lead to significant transport of mass and angular momentum in protoplanetary discs
\citep{2006ApJ...642.1140O,2006A&A...454..811P,2007A&A...462..193P}. In the present investigation we have used Eq.~(1) from
\citet{2006A&A...454..811P} to determine the encounter-induced relative disc-mass loss~$\Dmd$ of a star with mass~$M_1$ and disc
radius~$r_{\mathrm{d}}$ caused by a perturber with mass~$M_2$ and pericentric distance~$r_{\mathrm{p}}$:
\begin{equation}
  \begin{split}
    \Dmd &= \left( \frac{ M_2 }{ M_2 + 0.5 M_1 } \right)^{1.2} \log{ \left[ 2.8 \left( \frac{ r_{\mathrm{p}} }{
            r_{\mathrm{d}} } \right)^{0.1} \right] } \\
    &\times \exp\left\{-\sqrt{\frac{M_1}{M_2+0.5M_1}} \left[\left(\frac{r_{\rm{p}}}{r_{\rm{d}}}\right)^{3/2}-0.5\right]\right\} \,.
    \end{split}
    \label{eqn:ImprovedFit}
\end{equation}
Note that Eq.~\eqref{eqn:ImprovedFit} is valid for low relative disc masses only, $m_\mathrm{d} \lesssim 10^{-2} M_1$, implying negligible viscosity
and self-gravity. We point out that in this limit the estimated relative disc-mass loss $\Dmd$ is independent of the disc \emph{mass} but depends on
the disc \emph{radius}. However, it represents an upper limit because the parameter study was restricted to co-planar, prograde, parabolic encounters
that are the most perturbing. We note that Eq.~\eqref{eqn:ImprovedFit} has been derived for discs with a fixed surface density profile $\Sigma \propto
r^{-1}$. Recently, \citet{2012A&A...538A..10S} have shown that the shape of the disc-mass distribution has a significant impact on the quantity of the
disc-mass and angular momentum losses in star-disc encounters. Maximum losses are generally obtained for initially flat distributed disc material.

To account for the reduced disc-mass loss in hyperbolic encounters with eccentricity $\varepsilon > 1$, we use the fit function of the median relative
disc-mass loss, normalized to the parabolic case, as derived by \citet{2010A&A...509A..63O}:
\begin{equation}
  \label{eq:fit_function_mass_loss_vs_eccentricity}
  \begin{split}
    \Dmd(\varepsilon) &= \Dmd \\
    &\times \exp[-0.12(\varepsilon - 1)] \\
    &\times \{0.83 - 0.015(\varepsilon - 1) + 0.17 \exp[0.1(\varepsilon - 1)]\} \,.
  \end{split}
\end{equation}

Note that this fit is only a compromise: by using the median we underestimate the effect of strong perturbations (i.e. those induced by close
approaches of massive perturbers). However, as shown by \citet{2010A&A...509A..63O} this type of interactions is expected to occur rarely in stellar
systems as dense as the Arches cluster. 

The pure stellar dynamical simulations of a cluster environment have been carried out with the direct $N$-body integrator {\tt kira} as outlined in
Sec.~\ref{sec:numerical_method:cluster}. The dynamical data of each particle (time~$t$, mass~$M_1$, position~$\mathbf{r}$, and velocity~$\mathbf{v}$)
have been dumped every ten individual integration time steps to create a high-resolution temporal grid of particle phase space vectors. Finally, this
data set was parsed by a dedicated encounter tracking software to generate for each particle a record of interaction parameters with its strongest
perturber (time~$t$, particle mass~$M_1$, perturber mass~$M_2$, separation~$r$, and eccentricity~$\varepsilon$) at the beginning, the predicted
pericentre passage, and the end of the interaction. The algorithmic basics of the procedure are described in detail in
App.~\ref{app:encounter_tracking}.

In the remainder of the paper we will refer to \emph{normalized} disc-mass $\mdnorm$ and \emph{normalized} disc-mass loss $\Dmdnorm$. Normalizing the
disc-mass $\mdnorm \in [0,1]$ is justified when using Eqs.~\eqref{eqn:ImprovedFit} and~\eqref{eq:fit_function_mass_loss_vs_eccentricity} that are
independent of the absolute disc mass as outlined above. The normalized disc-mass loss $\Dmdnorm$ is defined as the \emph{absolute} loss of the
normalized disc-mass,
\begin{equation}
  \Dmdnorm \equiv \mdnorm \cdot \Dmd
  \label{eq:definition_normalized_disc_mass_loss} \,.
\end{equation}

In our approach a simplified prescription assigns stars into one of two distinct groups: if the accumulated normalized disc-mass loss $\Dmdnorm$
exceeds 0.9 (i.e. 90\,\% of the initial disc mass), stars are marked as ``disc-less''; otherwise they are termed ``star-disc systems''. According to
observations as outlined in Sec.~\ref{sec:introduction} we fix the typical disc diameter of a solar mass star to be 300\,AU and consider two different
model distributions of the initial disc radii~$r_\mathrm{d}$: i)~scaling with stellar mass~$M_1$ as $r_{\mathrm{d}} = 150\,\mathrm{AU} \sqrt{ M_1 /
  \Msun }$, which is equivalent to the assumption of an equal force at the disc boundary, and ii)~fixed disc radii $r_{\mathrm{d}} =
150\,\mathrm{AU}$. Whenever results are presented, we will specify which of these two distributions has been used. However, in the present study we
focus on the latter model as no obvious correlation between disc radius and stellar mass was found from observations so far
\citep[see][]{2005A&A...441..195V}. Note that we make no specific assumption about the distribution of disc masses. As outlined before the normalized
disc-mass loss $\Dmdnorm$ is independent of the absolute disc mass in the limit of low relative disc masses, $m_\mathrm{d} \lesssim 10^{-2} M_1$,
assumed here.

%

\section{Results of numerical simulations}

\label{sec:numerical_results}

For the two cluster models I and O a full mass distribution was generated according to the IMF published by \citet{2001MNRAS.322..231K}. Using these
two models we provide in the following a detailed view on the dynamical evolution of young stars and their circumstellar discs in the Arches cluster.

\subsection{Cluster dynamics}

\label{sec:numerical_results:cluster_dynamics}

The orbital model~O is a refined version of the isolated model~I that is evolved on a realistic orbit around the Galactic Centre in the Galactic tidal
field. It serves to analyze the evolution of the tidal features of the Arches cluster. However, unlike model~I, model~O is \emph{not} the best
representation of the internal dynamics of the present-day Arches cluster because it is not in a dynamical equilibrium state initially. The imposed
Galactic tidal field acts as an additional energy source such that model~O expands initially and re-bounces slightly before setting into an
equilibrium state.

\begin{figure}
  \centering
  \includegraphics[width=1.0\linewidth]{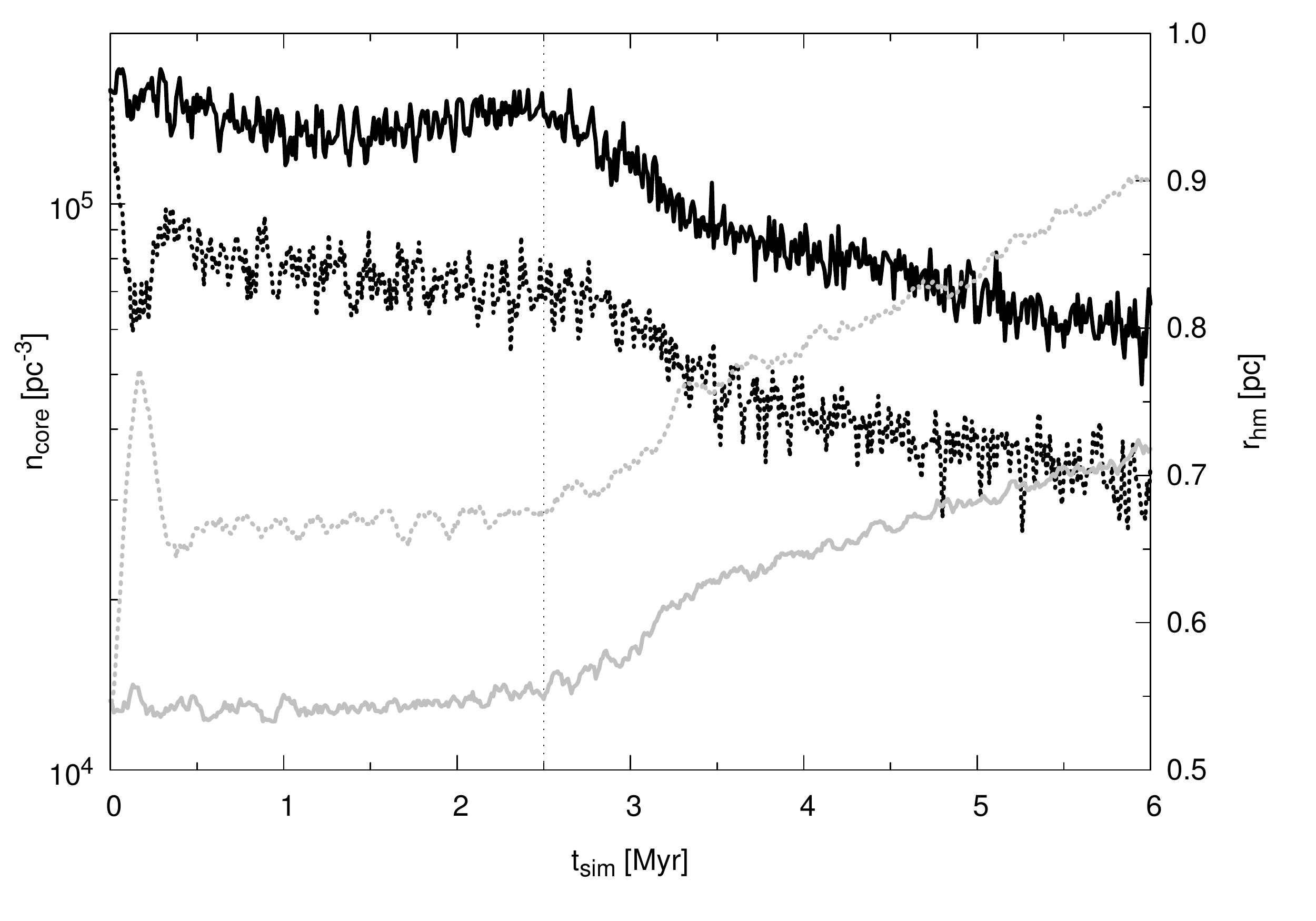}
  \caption{Comparison of the dynamical evolution of models I (solid) and O (dotted). The black and grey lines mark the core density (left logarithmic
    scale) and the half-mass radius (right linear scale), respectively.}
  \label{fig:dynamics__isolated_vs_orbit}
\end{figure}

We demonstrate this effect in Fig.~\ref{fig:dynamics__isolated_vs_orbit}. Initially, the core density~$n_{\mathrm{core}}$ of model~O (black dotted
line) drops rapidly by a factor of three, then rises slightly and afterwards remains roughly a factor of two below the value of model~I (solid black
line). The two lines are running parallel due to the logarithmic scale. An equivalent behaviour is seen for the half-mass radius (grey lines): the
ratio of the half-mass radii remains at about 1.3 (that is a factor of two in volume) after the re-bounce. Note the divergence of the two lines due to
the linear scale. Hence, after model~O has reached dynamical equilibrium at $\sim$0.3\,Myr the evolution of both models is dynamically equivalent over
the remaining 5.7\,Myr (or more than one hundred crossing times).

In particular do both models share the same characteristic local maximum of the core density~$n_{\mathrm{core}}$ at the present-day cluster age
$t_{\mathrm{AC}} \approx 2.5$\,Myr. This time~$t_{\mathrm{AC}}$ marks as well the minimum of the 1\,\% Lagrangian radius and the formation of a
massive binary. It is followed by a significant continuous expansion of the cluster as is evident from the evolution of the half-mass radius. These
features are characteristic signatures of core collapse \citep{1968MNRAS.138..495L,1983ApJ...271...11F}. We thus identify the expansion of the cluster
after $\sim$2.5\,Myr of dynamical evolution as its post-core-collapse phase.

\subsection{Encounter dynamics}

\label{sec:numerical_results:encounter_dynamics}

The main purpose of the present work is a realistic study of the encounter-induced disc-mass loss of stars in the Arches cluster by taking into
account the external effect of the Galactic tidal field. However, we have demonstrated in the previous section that the dynamical model with tidal
field, model~O, is too extended compared to model~I, our best-fitting isolated model of the observed properties of the Arches cluster. Consequently,
the encounter rate of star-disc systems in the underdense model~O is expected to be too low. Ignoring for a simplified dimensional argument the effect
of gravitational focusing, the frequency of encounters with distance~$r$ decreases with increasing radius~$R$ of a self-gravitating system,
\begin{equation}
  \label{eq:enc_freq}
  f_r = n \sigma \Sigma \propto R^{-3} R^{-1/2} r^2 = R^{-7/2} r^2 \,,
\end{equation}
where $n$, $\sigma$, and $\Sigma$ denote its density, velocity dispersion, and cross section with radius~$r$. The nearly constant size ratio of models
O and I over time, ${\cal{R}} = R^O/R^I \approx 1.3$, implies a roughly time-invariant ratio of encounter frequencies ${\cal{F}} = f_r^O / f_r^I
\propto {\cal{R}}^{-7/2} \approx 0.4$. However, Eq.~\eqref{eq:enc_freq} shows that we can compensate for this difference by reducing the encounter
distances~$r$ in the calculation of the disc-mass loss for model~O by a factor ${\cal{F}}^{1/2} \approx 0.6$. In fact, technically we have increased
the standard disc radius for model~O by ${\cal{F}}^{-1/2}$ to $r_{\mathrm{d}} = 250\,\mathrm{AU}$ (see Sec.~\ref{sec:numerical_method:encounters} for
reference) and found a very good agreement between the encounter histories of models O and I.

As will be shown later the only significant difference in the effect of encounters between models O and I that can not be corrected for is
gravitational focusing. Its strong dependence on stellar mass implies that the encounter-induced disc-mass loss of massive stars is reduced in
model~O. However, the general results for the entire cluster population are not much affected by the relatively rare interactions of high-mass stars
as discussed in Sec.~\ref{sec:numerical_method:encounters}.

\begin{figure}
  \centering
  \includegraphics[width=1.0\linewidth]{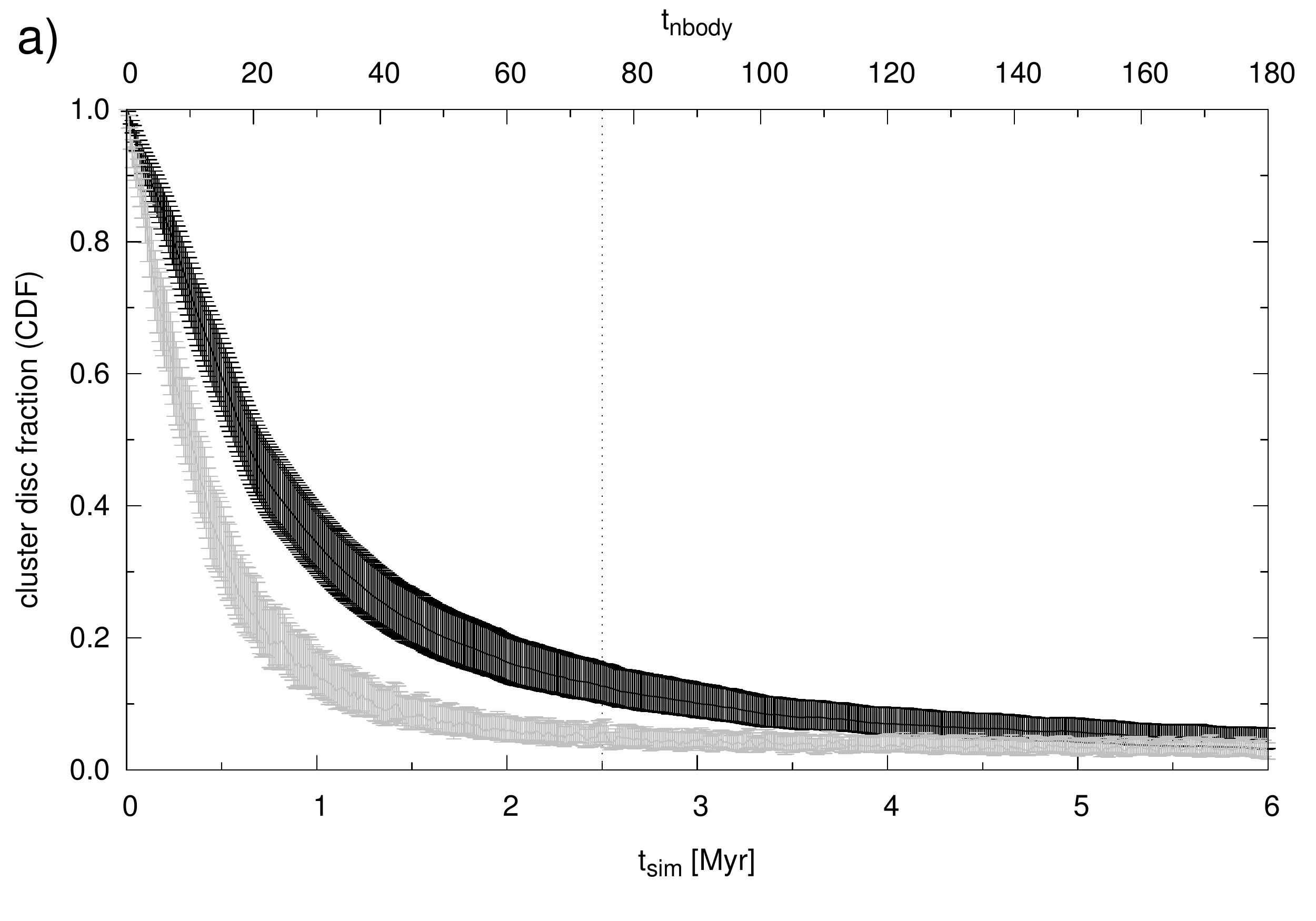}
  \includegraphics[width=1.0\linewidth]{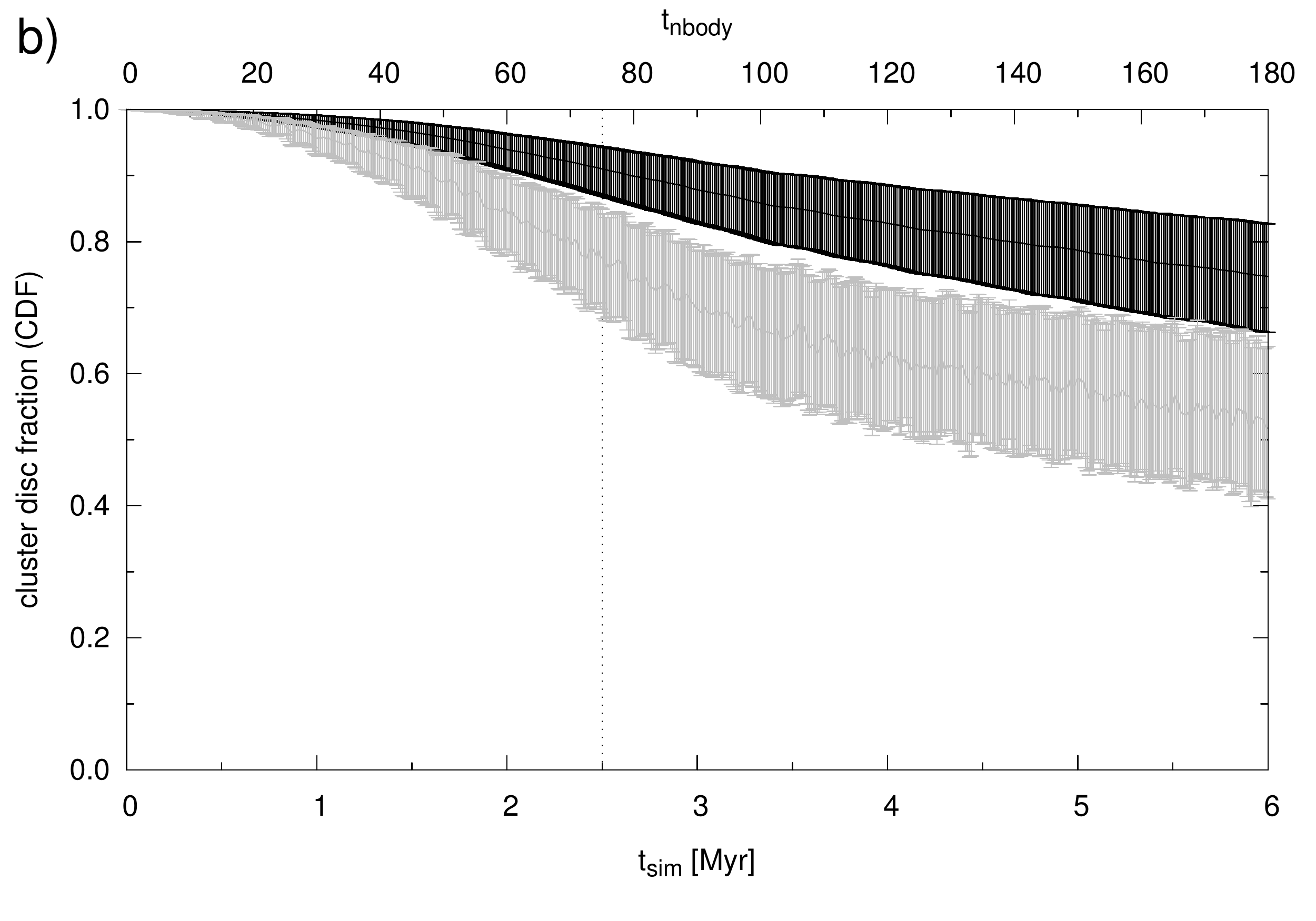}
  \caption{Time evolution of the encounter-driven cluster disc fraction (CDF) in two different volumes of model~O accounting for \textbf{a)}~parabolic
    encounters via Eq.~\eqref{eqn:ImprovedFit} and \textbf{b)}~eccentric encounters via Eq.~\eqref{eq:fit_function_mass_loss_vs_eccentricity}. Grey
    lines mark the core and black lines two times the half-mass radius at a given time. Constant initial disc radii have been assumed for this
    calculation (see Sec.~\ref{sec:numerical_method:encounters}). The vertical dotted line marks the estimated age of the Arches cluster. The time
    scale is shown in both physical (lower scale) and $N$-body units (upper scale).}
  \label{fig:cdf}
\end{figure}

The overall effect of the encounter-induced disc-mass loss in the Arches cluster is shown in Fig.~\ref{fig:cdf}. Here the temporal evolution of the
cluster disc fraction (CDF) is plotted. Note that stars with a normalized disc-mass loss of more than 0.9 are defined as disc-less. The upper plot has
been calculated using Eq.~\eqref{eqn:ImprovedFit}, restricted to parabolic encounters, while the bottom plot represents a more realistic scenario
considering the eccentricity of encounters via Eq.~\eqref{eq:fit_function_mass_loss_vs_eccentricity}. Here constant initial disc radii were assumed.

From Fig.~\ref{fig:cdf}a one would expect that more than 80\,\% of all discs are destroyed after 2.5\,Myr, and after 5\,Myr basically no discs would
have survived the repeating erosion by parabolic gravitational interactions. This scenario suggests that stellar encounters are potentially a very
efficient disc destruction mechanism in starburst clusters that could suppress planet formation entirely. However, unlike for intermediate-mass
clusters like the ONC the assumption of parabolic encounters is a strong simplification for dense stellar systems like the Arches cluster
\citep[see][]{2010A&A...509A..63O} and thus largely overestimates the disruptive effect on discs in this cluster. In hyperbolic encounters the
interaction time between the encounter partners is shorter and thus reduces the transport of energy and angular momentum \citep[see
e.g.][]{1994ApJ...424..292O}.

When we take the eccentricity of the stellar encounters into account (Fig.~\ref{fig:cdf}b) the effect on the disc-mass loss is much weaker, resulting
in the destruction of 10\,\% of the discs in the entire cluster (here defined as two times the half-mass radius) and roughly 30\,\% in the core up to
its present age. However, not only in the core but in the entire cluster the encounter-induced destruction of circumstellar discs does continue over
the entire simulation time and leads to a fractional disc destruction of 25\,\% in the entire cluster and 50\,\% in the core after 6\,Myr of dynamical
evolution.

The characteristic onset of expansion at $\sim2.5$\,Myr as identified in Fig.~\ref{fig:dynamics__isolated_vs_orbit} manifests itself in the kink of
the core disc fraction in the bottom panel of Fig.~\ref{fig:cdf} (grey line). The strong decrease in density reduces the stellar encounter rate and
thus the rate of disc destruction.

\begin{figure}
  \centering
  \includegraphics[width=1.0\linewidth]{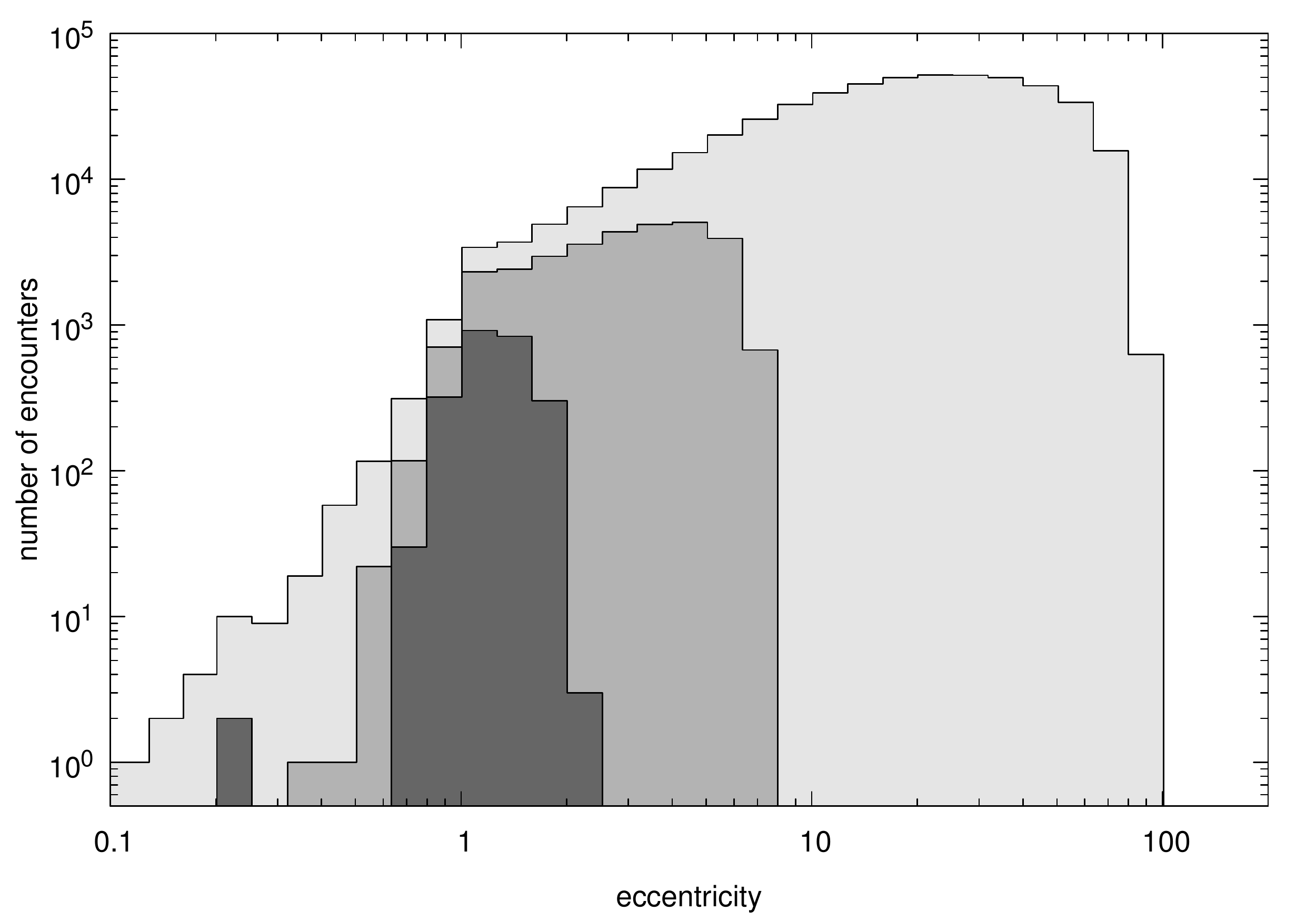}
  \caption{Number of encounters as a function of eccentricity for model~O. Constant initial disc radii have been assumed for this calculation (see
    Sec.~\ref{sec:numerical_method:encounters}). The different shaded regions mark all 516243 encounters (light grey), 31044 encounters causing more
    than 50\,\% (grey) and 2412 encounters with more than 90\,\% disc-mass loss (dark grey).}
  \label{fig:number_of_encounters__vs__eccentricity}
\end{figure}

We find that encounters of young star-disc systems in the Arches cluster can only destroy a minor fraction of all protoplanetary discs within a few
Myr. The encounter-induced disc-mass loss is drastically reduced by a large fraction of highly eccentric fly-bys. The reason is that hyperbolic
encounters typically result from chance encounters of low- and intermediate-mass stars while parabolic encounters are typically the outcome of
significant gravitational focusing of low-mass stars by massive stars \citep{2010A&A...509A..63O}. The corresponding encounter rates
$f_{\mathrm{hyp}}$ and $f_{\mathrm{par}}$ have very different scaling behaviour with the cluster density~$\rho$: $f_{\mathrm{hyp}} \propto \rho^{3/2}$
and $f_{\mathrm{par}} \propto \rho^{1/2}$. While the intermediate-mass ONC is dynamically balanced by hyperbolic and parabolic encounters \citep[see
Fig. 5 of][]{2010A&A...509A..63O}, in the roughly $20$ times denser Arches cluster the fraction of parabolic encounters is less than 10\,\%. Hence the
relative importance of gravitational focusing decreases very rapidly with increasing cluster density such that hyperbolic encounters become by far the
dominant type of stellar interactions in a starburst cluster. This is demonstrated for our model~O of the Arches cluster in
Fig.~\ref{fig:number_of_encounters__vs__eccentricity}. Here we plot the number of encounters as a function of eccentricity for three different
thresholds on the normalized disc-mass loss~$\Dmdnorm$ (marked by different grey scales). About 73\,\% of all encounters (light grey) have large
eccentricities $\varepsilon > 10$.

\begin{figure}
  \centering
  \includegraphics[width=1.0\linewidth]{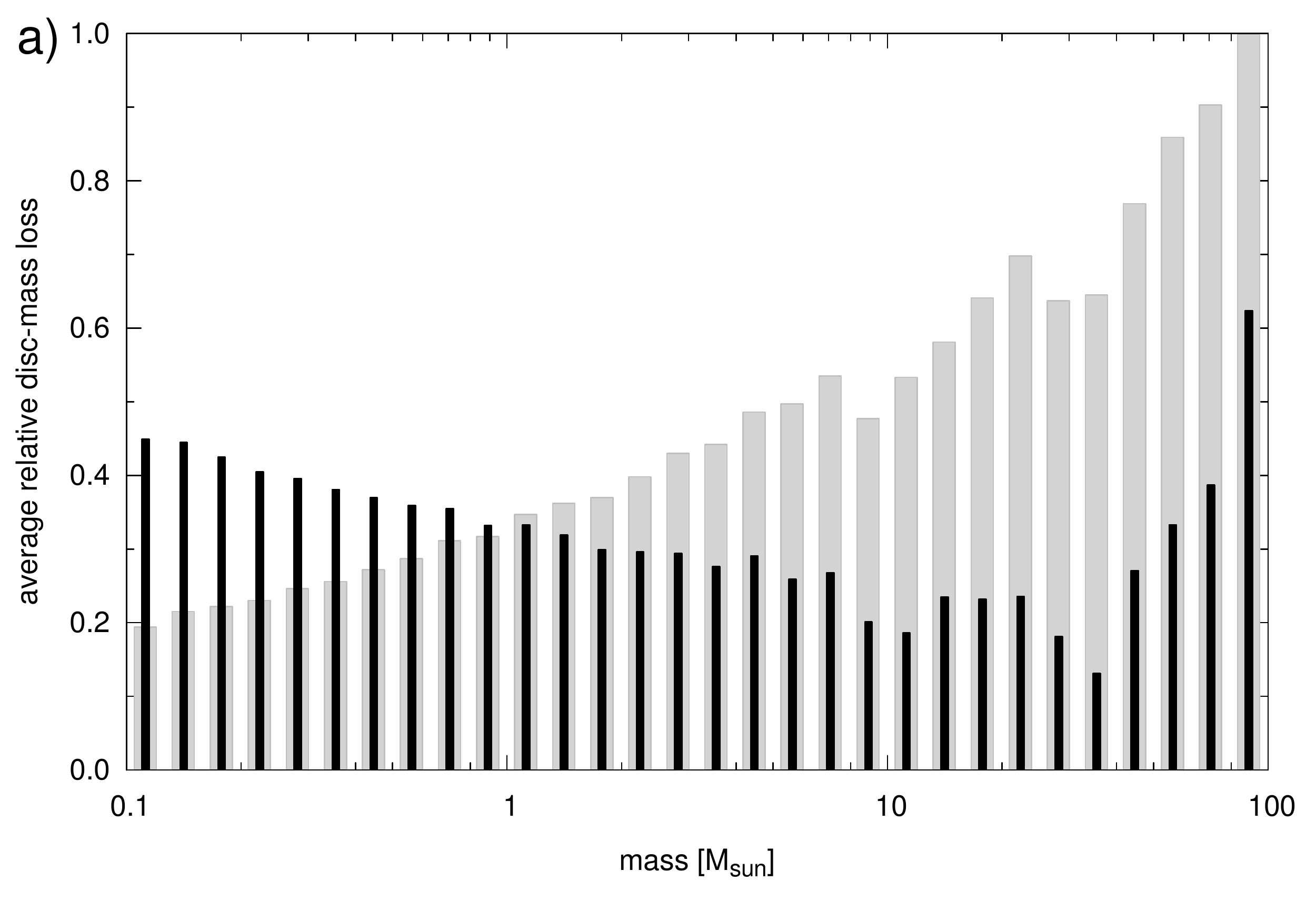}
  \includegraphics[width=1.0\linewidth]{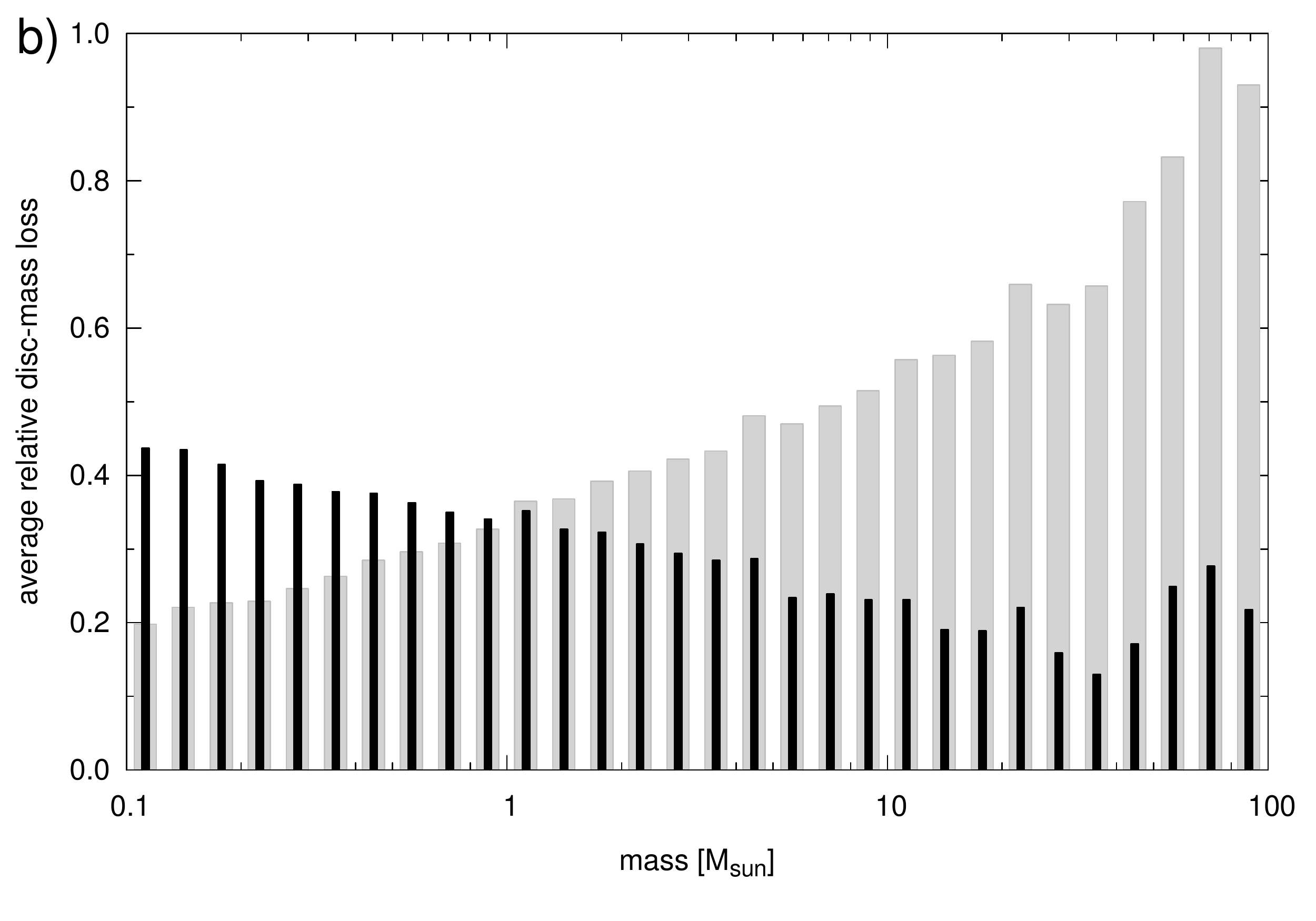}
  \caption{Average normalized disc-mass loss as a function of stellar mass after 2.5\,Myr of simulation time for \textbf{a)}~model~I and
    \textbf{b)}~model~O. The black and grey bars mark the scenarios using constant and scaled initial disc radii as outlined in
    Sec.~\ref{sec:numerical_method:encounters}, respectively. The underlying calculations take into account the eccentricity of the stellar encounters
    using Eq.~\eqref{eq:fit_function_mass_loss_vs_eccentricity}.}
  \label{fig:disc_mass_loss__vs__mass_bins}
\end{figure}


The following plots are thus based on calculations that take the eccentricity of the stellar encounters explicitly into account. In
Fig.~\ref{fig:disc_mass_loss__vs__mass_bins} we show the average normalized disc-mass loss as a function of stellar mass for two different initial
disc radius distributions (see Sec.~\ref{sec:numerical_method:encounters}) at the present age of the Arches cluster. The two plots show very similar
distributions of models I and O. As discussed at the beginning of this section the only significant difference occurs at the highest masses ($M >
40\,\Msun$). Their suppressed disc-mass loss for constant initial disc radii (black bars) in the two times sparser model~O
(Fig.~\ref{fig:disc_mass_loss__vs__mass_bins}b) is attributed to the less efficient gravitational focusing. So we will focus on the best-fitting
model~I for now (Fig.~\ref{fig:disc_mass_loss__vs__mass_bins}a).

If we assume that disc radii are positively correlated with stellar mass (grey bars) then stellar encounters would act entirely destructive on discs
of the most massive stars and least harmful at the lower mass end. However, in our preferred scenario of uncorrelated disc radii and stellar masses,
stars in the mass range $10-30\,\Msun$ would suffer least encounter-induced disc-mass loss while the lowest and highest mass stars would lose two to
three times more disc material, respectively. This result is analogous to simulations of the intermediate-mass ONC \citep[see][]{2006A&A...454..811P}
yet for the much denser and more massive Arches cluster the average disc-mass loss is lowest for roughly three times more massive stars (see their
Fig.~2).

The reason for this minimum is a mass-dependent bivariate encounter history: For low-mass stars the mass-loss occurs either through a few strong
parabolic encounter events with high-mass stars or frequent weak hyperbolic encounters with low- and intermediate-mass stars. High-mass stars lose
their discs via a steady nibbling by many gravitationally focused encounters with stars of lower mass \citep[][]{2010A&A...509A..63O}. For stars in
the intermediate mass regime both interaction channels are reduced such that encounter-induced disc-mass loss is least efficient.

\begin{figure}
  \centering
  \includegraphics[width=1.0\linewidth]{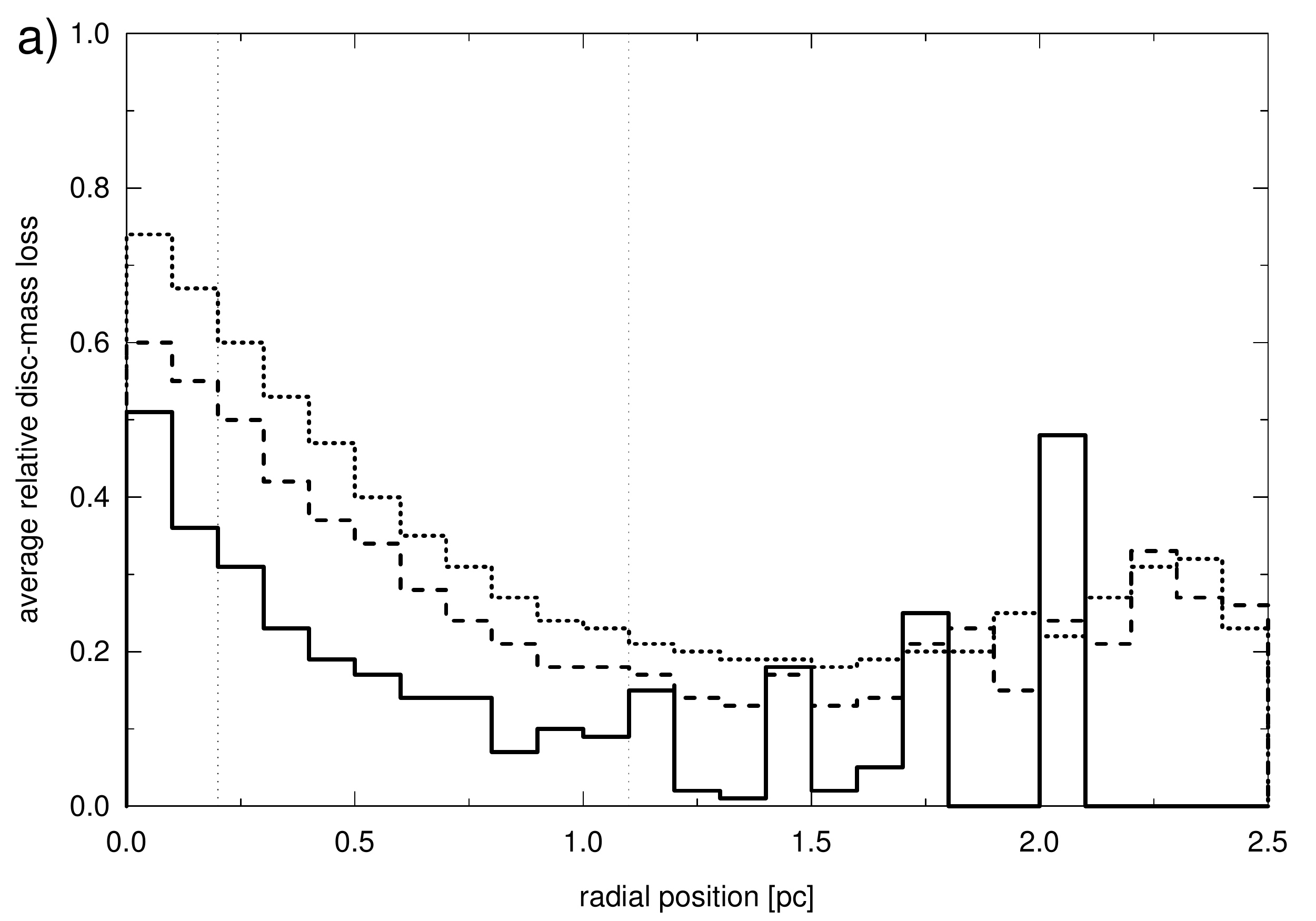}
  \includegraphics[width=1.0\linewidth]{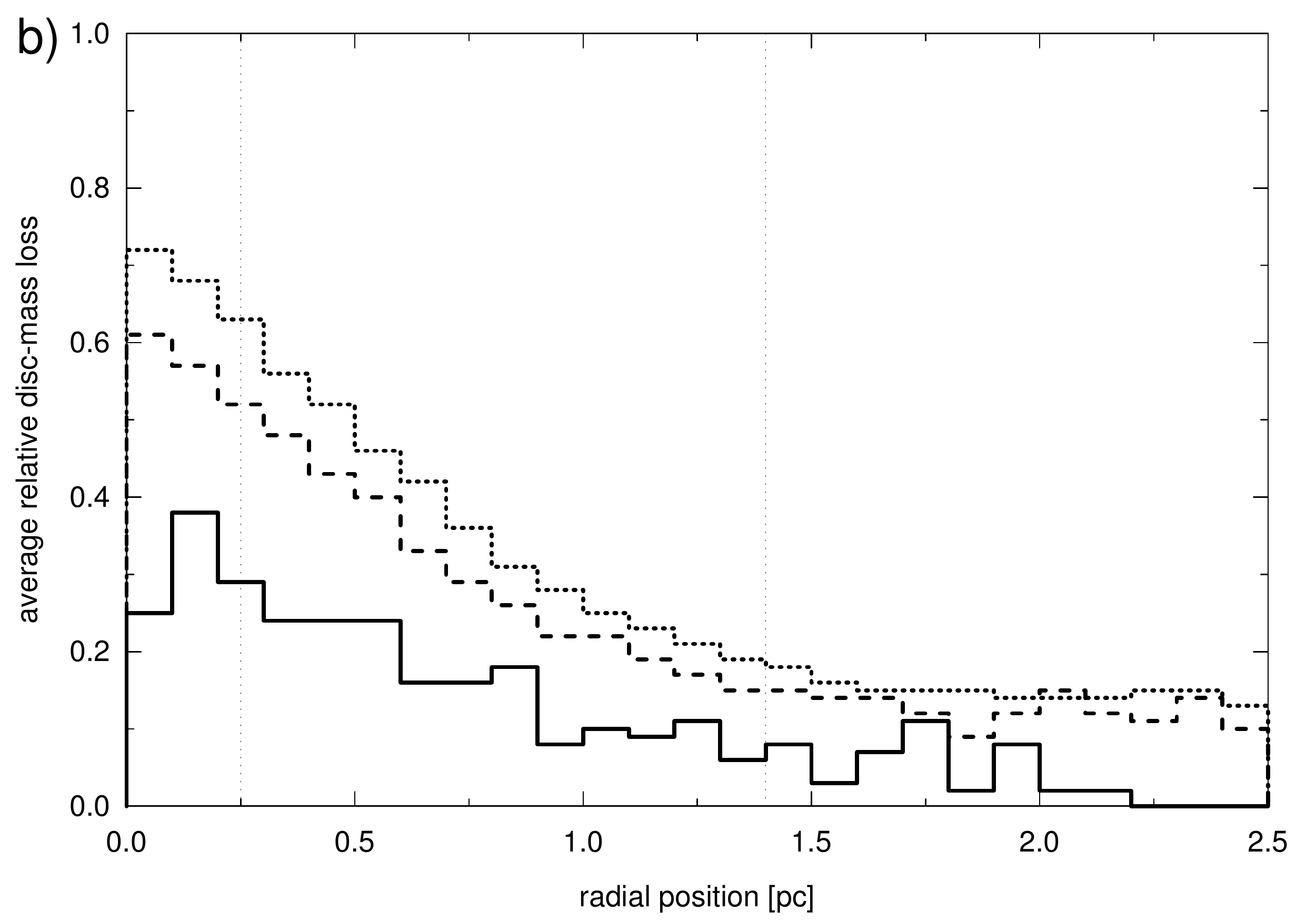}
  \caption{Average normalized disc-mass loss as a function of distance to the cluster centre after 2.5\,Myr of simulation time for \textbf{a)}~model~I
    and \textbf{b)}~model~O. The three lines mark three different mass groups of stars: $7-120\,\Msun$ (solid), $0.5-7\,\Msun$ (dashed), and
    $0.1-0.5\,\Msun$ (dotted). The underlying calculations take into account the eccentricity of the stellar encounters using
    Eq.~\eqref{eq:fit_function_mass_loss_vs_eccentricity}. Constant initial disc radii have been assumed for this calculation (see
    Sec.~\ref{sec:numerical_method:encounters}).}
  \label{fig:disc_mass_loss__vs__radius}
\end{figure}

The distribution of the resultant disc-mass loss with distance from the centre of the Arches cluster at its present age is presented in
Fig.~\ref{fig:disc_mass_loss__vs__radius}a for model~I. It confirms the expected decrease of the average normalized disc-mass loss with distance from
the cluster centre as a consequence of the decreasing local density. However, beyond roughly two half-mass radii -- the here used formal cluster
boundary -- the average normalized disc-mass loss rises again. This is the footprint of strong stellar interactions that occur preferentially in the
central region of the cluster. These interactions increase simultaneously the normalized disc-mass loss and probability for a rapid ejection of the
encountered star. The escape speed translates into distance from the cluster and explains its correlation with the normalized disc-mass loss outside
the cluster boundary. This can be seen best for the group of high-mass stars (solid line): the peaks in the histogram at large distances trace single
escapers with huge mass losses.

Unlike in model~I, the radial distribution of the average normalized disc-mass loss in model~O (Fig.~\ref{fig:disc_mass_loss__vs__radius}b) does not
show the characteristic increase outside of the cluster. This is because the sample is dominated by stars in the tidal tails of the cluster that do
not require strong interactions to become unbound from the cluster and thus mostly do not suffer increased disc-mass loss before escape. Note that the
lower disc-mass loss of the high-mass sample compared to model~I is again a consequence of the uncompensated gravitational focusing in model~O.

\begin{figure*}
  \centering
  \includegraphics[width=0.9\linewidth]{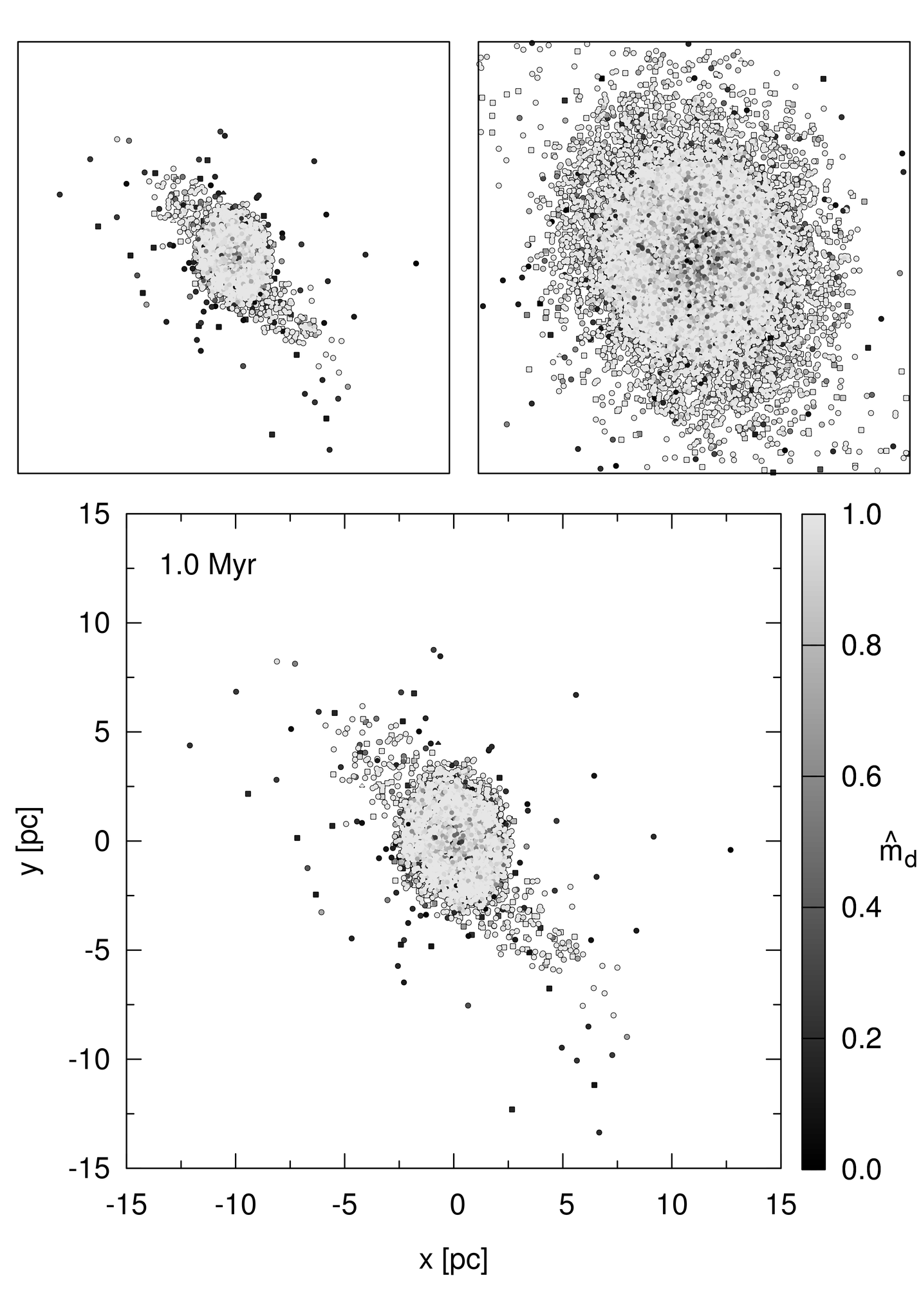}
  \caption{Snapshots of the stellar population with encounter-induced disc evolution in the Arches cluster for the orbital model~O after 1\,Myr of
    dynamical evolution. The bottom panel shows the full extension of the cluster, the two smaller top panels are zooms of 15\,pc box half-width
    (left) and 3.5\,pc box half-width (right), respectively. The grey scale represents the normalized disc mass~$\mdnorm$; the origin of the
    coordinate system marks the cluster centre-of-mass. The symbols represent three different stellar mass ranges: $7-120\,\Msun$ (triangle),
    $0.5-7\,\Msun$ (rectangle), and $0.1-0.5\,\Msun$ (circle). The underlying calculations take into account the eccentricity of the stellar
    encounters using Eq.~\eqref{eq:fit_function_mass_loss_vs_eccentricity}. Constant initial disc radii have been assumed for this calculation (see
    Sec.~\ref{sec:numerical_method:encounters}).}
  \label{fig:cluster__disc_mass__1_Myr}
\end{figure*}
\begin{figure*}
  \centering
  \includegraphics[width=0.9\linewidth]{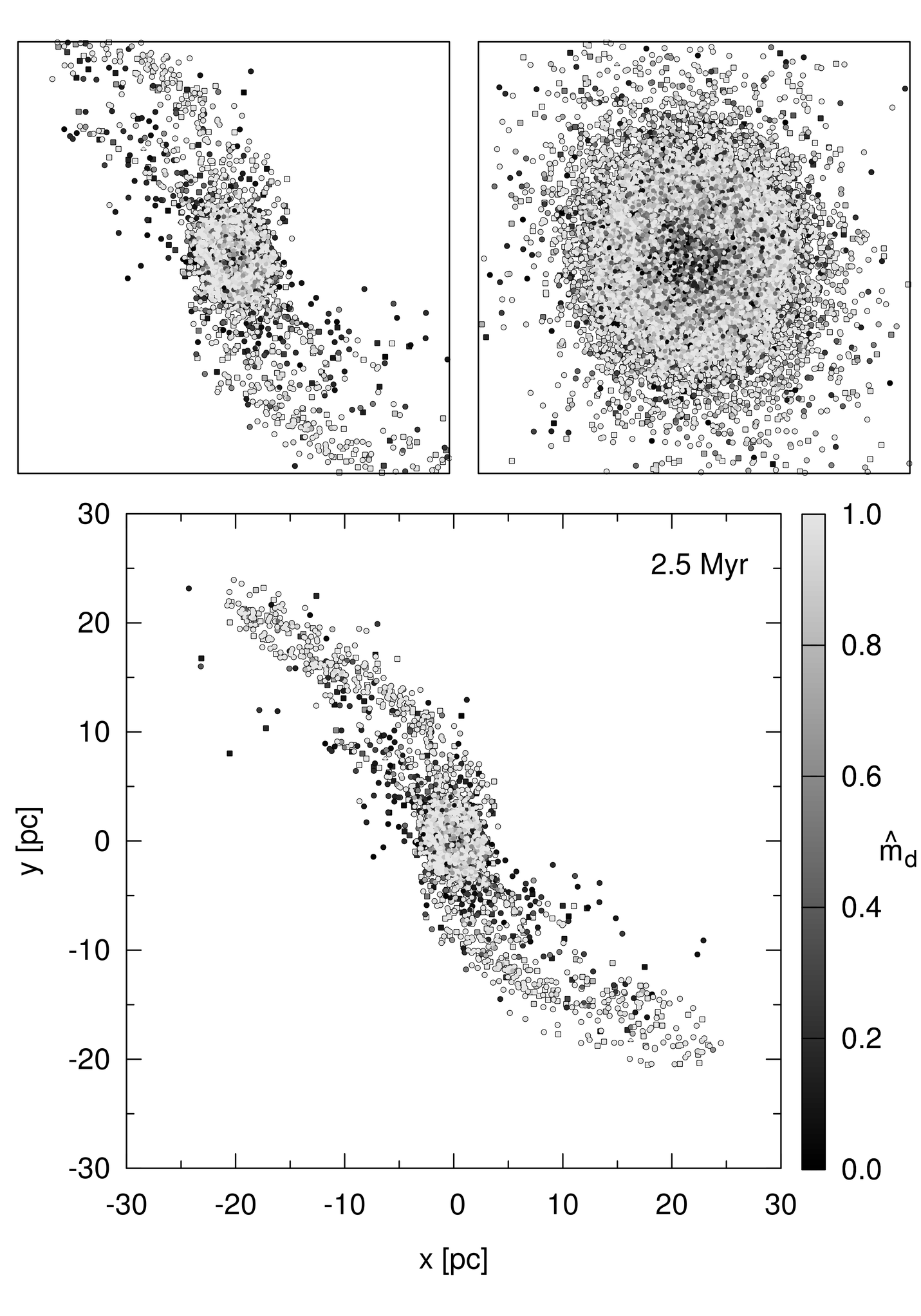}
  \caption{Same as Fig.~\ref{fig:cluster__disc_mass__1_Myr} after 2.5\,Myr of dynamical evolution.}
  \label{fig:cluster__disc_mass__2.5_Myr}
\end{figure*}
\begin{figure*}
  \centering
  \includegraphics[width=0.9\linewidth]{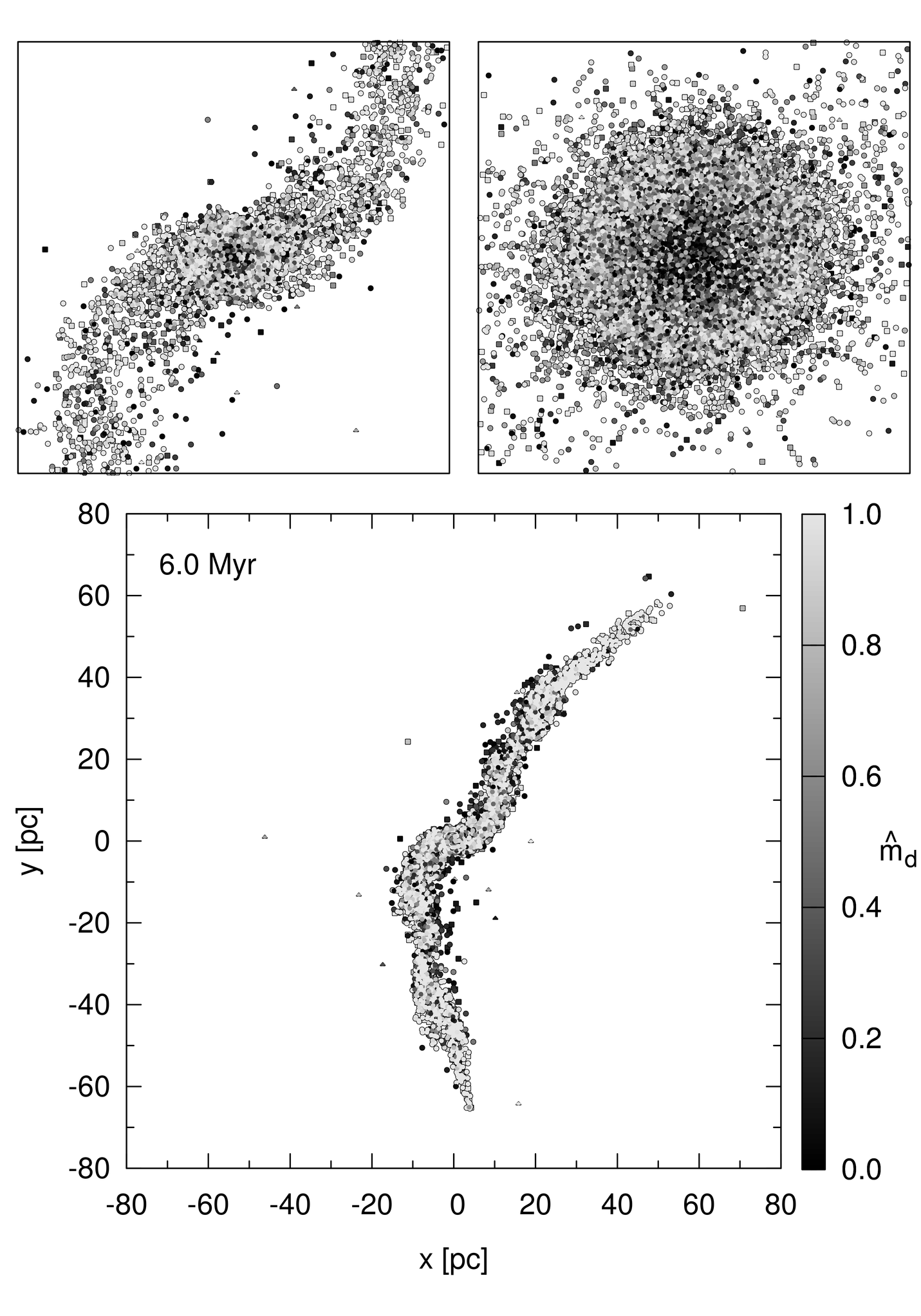}
  \caption{Same as Fig.~\ref{fig:cluster__disc_mass__1_Myr} after 6\,Myr of dynamical evolution.}
  \label{fig:cluster__disc_mass__6_Myr}
\end{figure*}
In the previous plots we have shown the evolution of the averaged disc-mass loss binned over different quantities. Next we present a more detailed
picture by showing the evolution of the spatial distribution of the disc-mass loss on a star-by-star basis at 1\,Myr, 2.5\,Myr, and 6\,Myr in
Figs.~\ref{fig:cluster__disc_mass__1_Myr} to \ref{fig:cluster__disc_mass__6_Myr}. The bottom panel shows the full extension of the cluster, the two
smaller top panels are zooms of 15\,pc box half-width (left) and 3.5\,pc box half-width (right), respectively. The grey scale represents the
normalized disc mass~$\mdnorm$. In agreement with Figs.~\ref{fig:cdf} and \ref{fig:disc_mass_loss__vs__radius} the fraction of disc-poor stars (dark
symbols) increases towards the cluster centre and with time. However, the most prominent features are the growing tidal tails formed by stars escaping
from the cluster in the tidal field imposed by the Galactic potential. By the end of the calculation the entire structure extends over an arc of
nearly 150\,pc length (Fig.~\ref{fig:cluster__disc_mass__6_Myr}).

\begin{figure}
  \centering
  \includegraphics[width=1.0\linewidth]{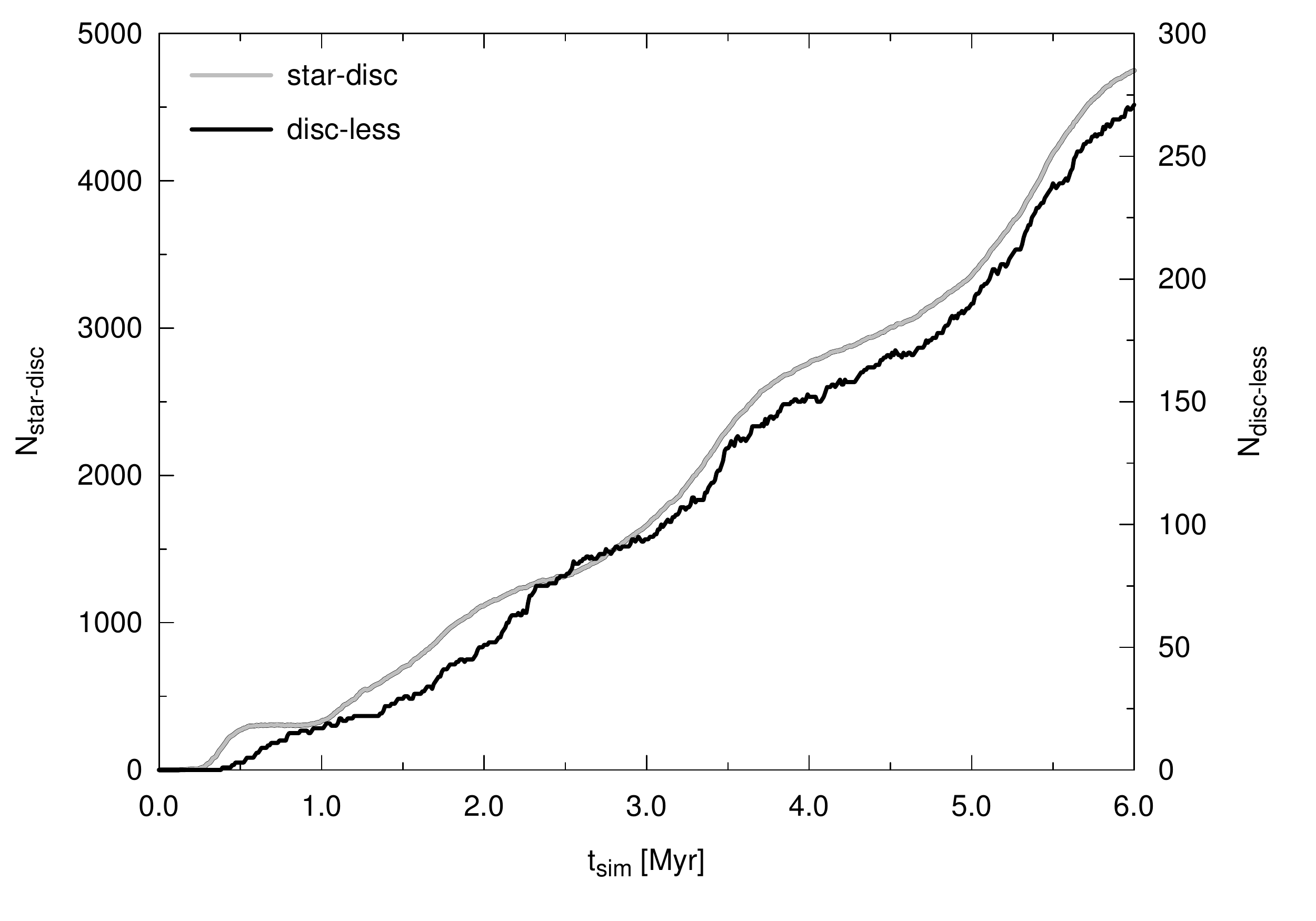}
  \caption{Time evolution of the number of star-disc systems (grey line, left scale) and disc-less stars (black line, right scale) in the tidal tails
    of model~O. The underlying calculations take into account the eccentricity of the stellar encounters using
    Eq.~\eqref{eq:fit_function_mass_loss_vs_eccentricity}. Constant initial disc radii have been assumed for this calculation (see
    Sec.~\ref{sec:numerical_method:encounters}).}
  \label{fig:tidal_tails__population}
\end{figure}

However, we show in Fig.~\ref{fig:tidal_tails__population} that despite its proximity to the deep gravitational potential of the Galactic centre only
about 5000 stars or 7\,\% of the entire population are part of the tidal tails after 6\,Myr of dynamical evolution. Here we define stars outside a
distance of 3.5\,pc to the cluster centre as escapers or equivalently as tidal tail members. The fraction of disc-less stars in the tidal tails
remains roughly constant over time at about 5\,\% and amounts to 267 stars at 6\,Myr. The escape rate is not constant but has periodical local maxima
that coincide with the cluster's pericentre passages. It's shape is similar for star-disc systems and disc-less stars.

\begin{figure}
  \centering
  \includegraphics[width=1.0\linewidth]{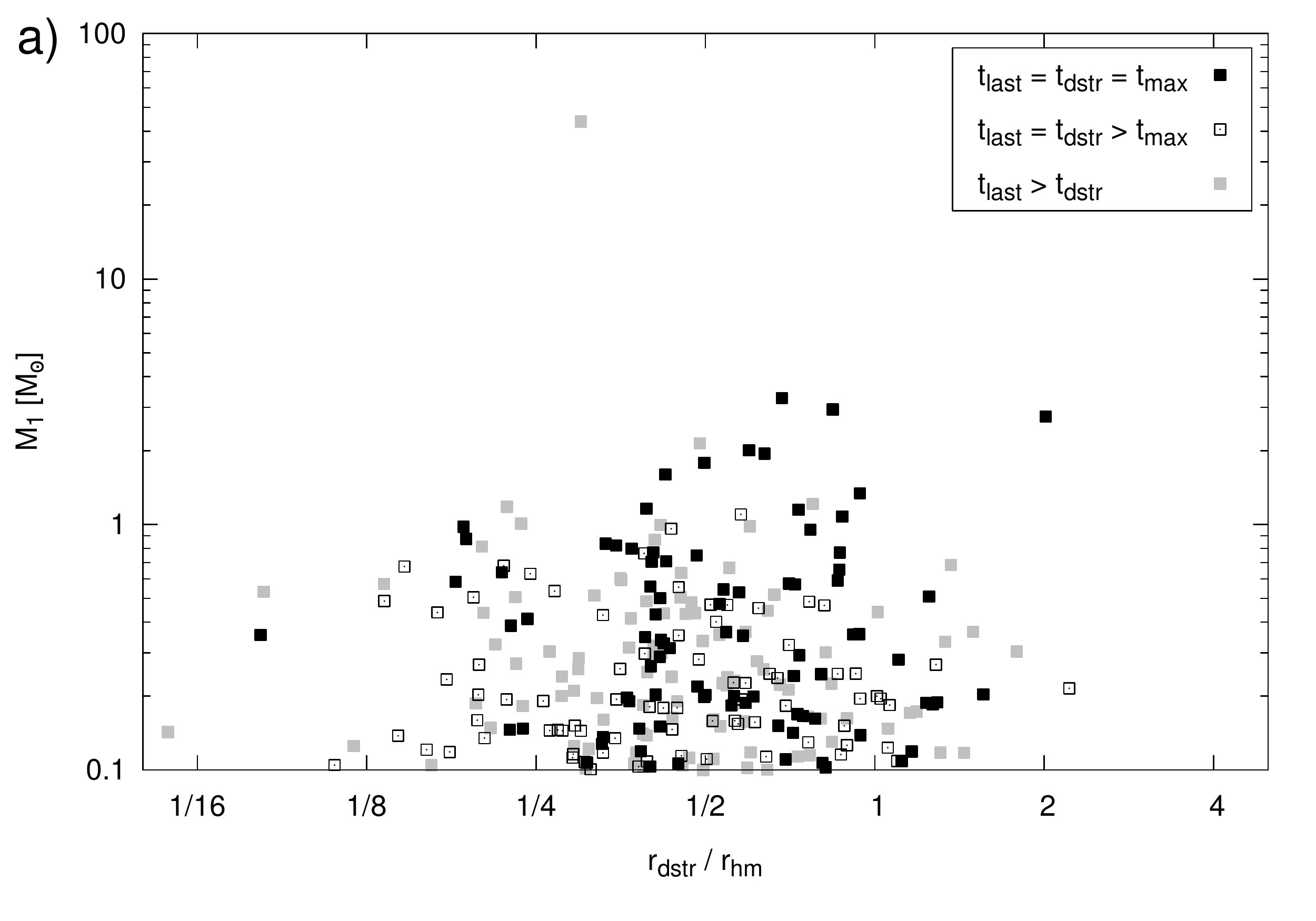}
  \includegraphics[width=1.0\linewidth]{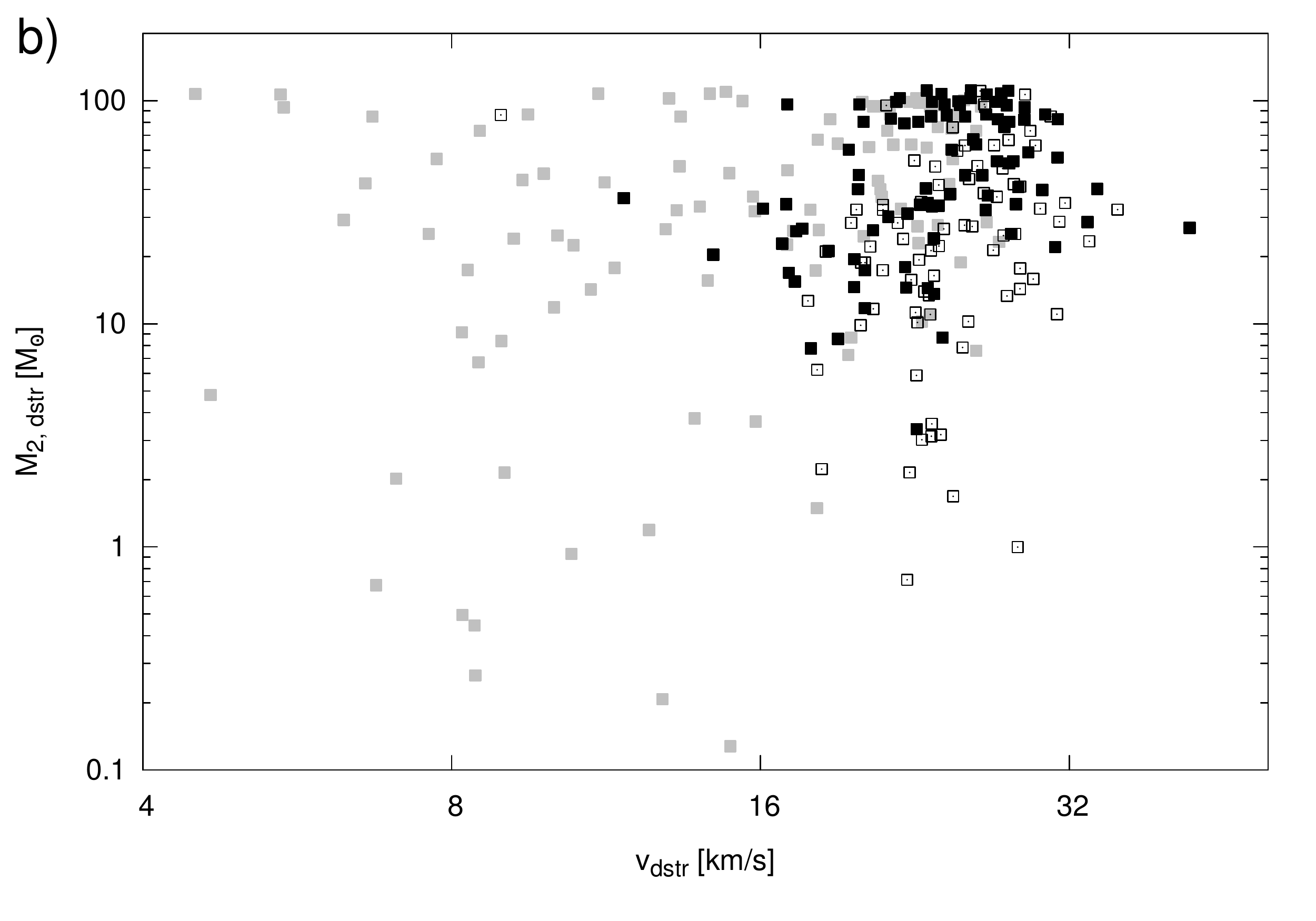}
  \includegraphics[width=1.0\linewidth]{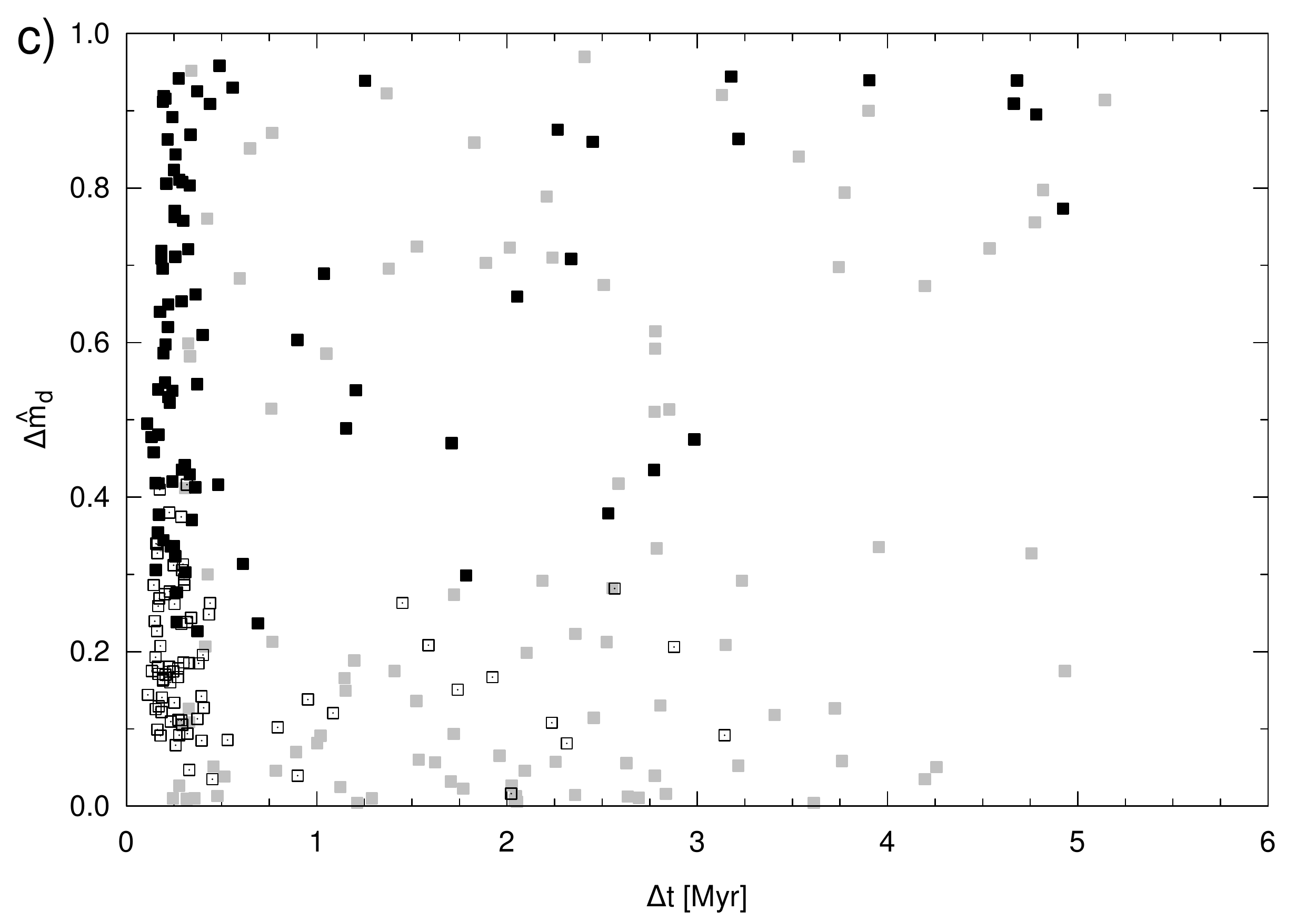}
  \caption{Distributions of characteristic dynamical quantities of disc-less tidal tail stars from the three sets $\tailstarsA$ (black filled
    squares), $\tailstarsB$ (black open squares), and $\tailstarsC$ (grey filled squares). The panels show \textbf{a)}~stellar mass $M_1$ vs. relative
    distance to the cluster centre after disc destruction $r_\mathrm{dstr}$ in terms of cluster half-mass radii $r_\mathrm{rh}$, \textbf{b)}~perturber
    mass $M_\mathrm{2,dstr}$ that caused disc destruction vs. stellar velocity~$v_\mathrm{dstr}$ after disc destruction, and \textbf{c)}~normalized
    disc-mass loss $\Dmdnorm$ vs. time delay~$\Delta{t}$ between disc destruction and escape.}
  \label{fig:stellar_encounter_types}
\end{figure}

For a deeper understanding of the dynamical history of the disc-less tidal tail stars in the following we investigate the temporal occurrence of three
specific types of encounters for each star. The events are referenced to the time of periastron passage of a perturber. At time $t_\mathrm{max}$ a
star experiences the largest normalized disc-mass loss~$\Dmdnorm$ due to an encounter. Note that for two events with equal \emph{relative} disc-mass
loss $\Dmd$ the first one causes a larger \emph{normalized} disc-mass loss $\Dmdnorm$, as is evident from
Eq.~\eqref{eq:definition_normalized_disc_mass_loss}. Hence $t_\mathrm{max}$ marks the event with largest absolute loss of the normalized disc mass
$\mdnorm$, which is not necessarily coincident with the strongest perturbation of the disc. An event at $t_\mathrm{dstr}$ marks encounter-induced disc
destruction. According to our criterion given in Section~\ref{sec:numerical_method:encounters} this is the first encounter that reduces the normalized
disc mass $\mdnorm$ below 0.1. The encountered star is considered to be disc-less subsequently. Finally, the time $t_\mathrm{last}$ marks the last
encounter before escape from the cluster. Note that this does not necessarily imply the event induced an ejection from the cluster. It is possible for
a star to be accelerated above escape speed in a previous encounter and to experience only a weak last encounter on its way out of the
cluster. However, the low probability for such an event implies that in most cases the encounter at $t_\mathrm{last}$ is energetic enough to unbind
the star from the cluster.

In the following we use the coincidence of these three dynamical events to split the entire set of 267 disc-less tidal tail stars at the end of the
simulation, $\tailstars$, into four disjoint subsets (marked in Fig.~\ref{fig:stellar_encounter_types} by the symbols given in parentheses):
\begin{equation*}
  \begin{aligned}
    \tailstarsA &= \{ x \in \tailstars \mid t_\mathrm{last} = t_\mathrm{dstr} = t_\mathrm{max} \} &\text{(black filled)}\,,\\
    \tailstarsB &= \{ x \in \tailstars \mid t_\mathrm{last} = t_\mathrm{dstr} > t_\mathrm{max} \} &\text{(black open)}\,,\\
    \tailstarsC &= \{ x \in \tailstars \mid t_\mathrm{last} > t_\mathrm{dstr} \}                  &\text{(grey filled)}\,,\;\\
    \tailstarsD &= \{ x \in \tailstars \mid t_\mathrm{last} < t_\mathrm{dstr} \} \,.\\
  \end{aligned}
\end{equation*}
Subset $\tailstarsA$ contains all the stars that lost their disc in the most perturbing encounter, being simultaneously ejected from the cluster. This
implies highly energetic interactions. The difference to subset $\tailstarsB$ is that disc destruction for those stars occurred in a subsequent weaker
event before escape from the cluster. However, here the last encounter was probably still quite energetic as simultaneous disc destruction and
ejection are less probable for a weak interaction. In the third set $\tailstarsC$ we group stars with disc destruction before the last encounter. The
dynamical history of these stars is potentially quite diverse. The disc might be destroyed in a rather strong encounter that does not immediately lead
to ejection from the cluster. This is either because the encounter is not sufficiently energetic to unbind a star residing deep in the cluster
potential or the potential escaper is back-scattered in a subsequent encounter. Alternatively, disc destruction is driven by a series of weak
encounters. Eventually the disc-less star either evaporates out of the cluster or is ejected via an energetic encounter. Membership in the last subset
$\tailstarsD$ is restricted to stars with disc destruction after escape from the cluster. The cardinality of the four sets is $|\tailstarsA| = 88$,
$|\tailstarsB| = 79$, $|\tailstarsC| = 100$, and $|\tailstarsD| = 0$.

We expect the energetic encounters of stars in set~$\tailstarsA$ to involve preferentially a massive perturber and rapid ejection from the cluster
with high velocity. We expect a similar scenario for members of~$\tailstarsB$ but a larger fraction of lower perturber masses due to the lower
normalized disc-mass loss before ejection. In contrast, most members of set~$\tailstarsC$ probably do not escape directly after disc destruction but
remain bound to the cluster for some time. We expect disc destruction by perturbers of even lower mass without significant acceleration. An
interesting consequence of the emptiness of~$\tailstarsD$ is that disc destruction does not occur in the tidal tails within the first 6\,Myr of
cluster evolution. In other words all disc-less stars in the tidal tails have lost their disc before escape.

In Fig.~\ref{fig:stellar_encounter_types} we present the dynamical properties of the disc-less tidal tail stars in the three non-empty
subsets. Panel~a) shows that disc-less tidal tail stars have mostly low mass~$M_1$ (with a median of 0.23\,$\Msun$) and lose their disc at a small
distance $r_\mathrm{dstr}$ to the cluster centre (with a median of $\nicefrac{1}{2} R_\mathrm{hm} \approx 0.4\,\pc$). Only one very massive star with
$M_1 > 10\,\Msun$ becomes a disc-less tidal tail star as a member of~$\tailstarsC$. However, the distributions of $M_1$ and $r_\mathrm{dstr}$ do not
differ significantly between the three subsets. In contrast, and as expected, panel~b) clearly shows different distributions of perturber masses
$M_\mathrm{2,dstr}$ that caused disc destruction and stellar velocities~$v_\mathrm{dstr}$ after disc destruction. The lowest perturber masses for the
three subsets $\tailstarsA$, $\tailstarsB$, and $\tailstarsC$ are $3.4\,\Msun$, $0.72\,\Msun$, and $0.13\,\Msun$, respectively. The velocity
distributions differ strongly between set~$\tailstarsC$ and the union set $\tailstarsAB \equiv \tailstarsA \cup \tailstarsB$. While half of the stars
in $\tailstarsC$ have $v_\mathrm{dstr} < 16\,\kms$, this is the case for only 3 of 167 stars in~$\tailstarsAB$. So formally most of these latter stars
are escapers with $v > 2\sigma^O_{3D}$, where $\sigma^O_{3D} \approx 10\,\kms$ is the three-dimensional velocity dispersion of model~O. We thus
conclude that nearly two thirds of the disc-less tidal tail stars lose their disc in a strong scattering event with a massive perturber. However, the
distribution of the normalized disc-mass loss $\Dmdnorm$ as a function of the time delay~$\Delta{t}$ between disc destruction and escape in panel~c)
shows that not all of these formal escapers are immediately expelled from the cluster. A fraction of 39/167 stars from the union set~$\tailstarsAB$
remains in the cluster for more than 1\,Myr, even after very strong encounters with $\Dmdnorm > 0.9$. So even the strongest perturbations do not
necessarily lead to the ejection of stars from the cluster. On the other hand does a small fraction of 17/100 stars from set~$\tailstarsC$ escape
within a few 0.1\,Myr. In total after 6\,Myr 145 of 267 disc-less stars have escaped within 0.5\,Myr to become part of the tidal tails. We find no
correlation between the normalized disc-mass loss causing disc destruction and the subsequent time to escape from the cluster.

\begin{figure}
  \centering
  \includegraphics[width=1.0\linewidth]{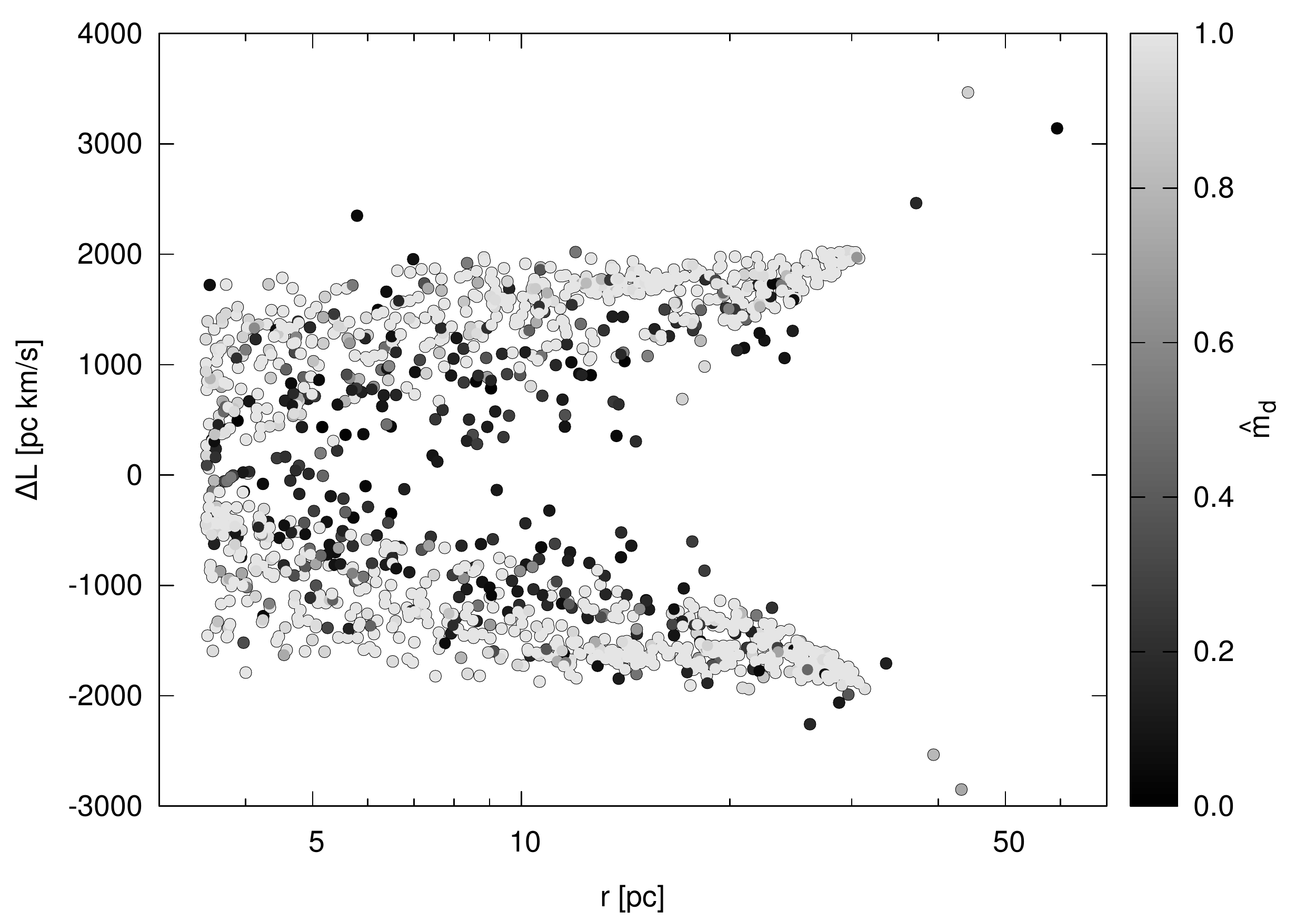}
  \caption{Relative specific angular momentum~$\Delta{L}$ of stars in the tidal tails of model~O with respect to the cluster centre-of-mass as a
    function of their distance from the cluster centre at 2.5\,Myr. The grey scale represents the normalized disc mass~$\mdnorm$. The underlying
    calculations take into account the eccentricity of the stellar encounters using Eq.~\eqref{eq:fit_function_mass_loss_vs_eccentricity}. Constant
    initial disc radii have been assumed for this calculation (see Sec.~\ref{sec:numerical_method:encounters}).}
  \label{fig:tidal_diagnostics}
\end{figure}

In agreement with the rapid ejection of more than half of the disc-less stars, Figs.~\ref{fig:cluster__disc_mass__1_Myr} and
\ref{fig:cluster__disc_mass__2.5_Myr} show that there is a clear difference in the spatial distribution of star-disc systems and disc-less stars in
the tidal tails. The disc-poor stars are preferentially located at large distances perpendicular to the tidal arms (i.e. along the orbital motion),
and close to the cluster along the direction of the tidal arms (i.e. radially to the Galactic Centre). This is a consequence of strong interactions
that remove disc material and accelerate stars sufficiently to overcome the deep cluster potential and escape in arbitrary directions -- not
preferentially via the Lagrange points. The specific angular momentum~$L$ of these stars is less altered than that of stars escaping radially through
the Lagrange points. We demonstrate this effect in Fig.~\ref{fig:tidal_diagnostics} by plotting the relative specific angular momentum of tidal tail
stars $\Delta{L} = L - L_\mathrm{C}$, where $L_\mathrm{C}$ is the specific angular momentum of the cluster centre-of-mass. Clearly, the modulus
$|\Delta{L}|$ is lower for disc-poor stars and thus explains their smaller radial offset by Eq.~\eqref{eq:just09}.

As a consequence of the different dynamical history of tidal tail stars we observe the formation of two classes of tidal tails in the present-day
cluster (Fig.~\ref{fig:cluster__disc_mass__2.5_Myr}): a shorter ``disc-poor'' tidal arm pair populated by stars expelled from the cluster and
suffering huge disc-mass loss (\mbox{$\gtrsim\,80\,\%$}) and the ``classical'' arms that are formed by stars evaporating though the Lagrange
points. This is how -- according to our dynamical model -- the Arches cluster would appear at present.

With time the growing number of stars that are tidally stripped from the cluster mix with the disc-poor escapers and the two types of arms merge into
one large structure (Fig.~\ref{fig:cluster__disc_mass__6_Myr}). However, the disc-poor stars still populate two distinct regions: the edge close to
the Galactic centre in the trailing arm and the far edge in the leading arm.

%

\section{Summary and Discussion}

\label{sec:conclusions}

We have investigated the effect of stellar encounters on the evolution of protoplanetry discs using a numerical model of the Arches cluster, one of
the most extreme Galactic star clusters observed so far.

Our numerical method is based on a hybrid approach where we combine simulations of pure star cluster dynamics and a parameter study of isolated
star-disc encounters.\footnote{``star-disc encounters'' refer to the interaction of a star-disc system and a disc-less star and serve as a numerical
  simplification of the disc-disc encounters that occur in reality.} This approach involves necessarily some assumptions and simplifications that
potentially mildly overestimate the effect of the encounter-induced perturbations of circumstellar material. Nonetheless, our work represents the most
detailed study of the effect of stellar dynamics on the evolution of protoplanetary discs in a starburst cluster so far.

We find that stellar interactions in the Arches cluster are dominated by hyperbolic fly-bys, with more than 70\,\% having eccentricities $\varepsilon
> 10$. This is much different in sparser clusters where nearly parabolic encounters are by far most frequent \citep{2010A&A...509A..63O}. According to
Eq.~\eqref{eq:fit_function_mass_loss_vs_eccentricity} the normalized disc-mass loss in hyperbolic interactions with $\varepsilon > 10$ is reduced by a
factor $<$\,0.38 compared to the parabolic case. Consequently, encounters drive the destruction of only about one third of the stellar discs in the
cluster core until its present-day age of $\sim$2.5\,Myr. The process continues with reduced efficiency such that after 6\,Myr half of the core
population becomes disc-less. The extended period of disc destruction in a starburst cluster is again in stark contrast to sparser clusters in which
it ceases after roughly 1\,Myr due to significant cluster expansion \citep{2010A&A...509A..63O}. Note that by neglecting primordial binaries in our
models we underestimate the efficiency of disc destruction for three reasons: (i) the larger number of stars increases the probability for encounters,
(ii) binaries increase the probability for strong few-body interactions, and (iii) discs are truncated in tight binaries.

The inclusion of a Galactic tidal field in our model of the Arches cluster induces the growth of an extended pair of tidal tails. The fraction of
disc-less stars in the tidal tails is nearly constant over time at $\sim$5\,\% and hence much lower than in the cluster at ages later than 1\,Myr.
Only about half of the disc-less tidal tail stars have escaped slowly via the Lagrange points while the other half has been ejected at high speed from
the cluster centre directly after disc destruction. These stars are the dominant population of a weaker and shorter second pair of ``disc-poor'' tidal
arms at the present age of the cluster. With time the disc-poor tails mix with the dominant ``classical'' tails and form new features: a concentration
of disc-less stars at the inner edge of the trailing arm and the outer edge of the leading arm.

In fact, we expect that dynamical features arising from strong encounters -- such as the disc-poor tidal arms -- could be even more prominent than
shown by our model. For one, as outlined in Section~\ref{sec:numerical_results:cluster_dynamics}, our model of the Arches cluster in the Galactic
tidal field is slightly underdense such that close encounters are underrepresented in the simulation. Second, we model a population of single stars
only and ignore the dynamically important effect of binaries. Assuming the binary fraction in starburst clusters is potentially as high as in
clustered star forming regions of lower mass \citep[$\sim$50\,\%: e.g.][]{2003ApJ...583..358B,2006A&A...458..461K}, the probability for strong
few-body encounters would increase significantly \citep{1975AJ.....80..809H,1975MNRAS.173..729H,1986Ap&SS.124..217A}, boosting the production of
high-speed escapers and disc-less stars.

However, one could argue that taking into account additional disc destruction mechanisms that are independent of cluster dynamics could mask the
difference between the classical and disc-poor tidal arms. We discuss here shortly the potential contribution from two other important processes. For
one, grain growth \citep{2003A&A...400L..21V,2006ApJ...644L..71S,2008A&A...480..859B,2008ARA&A..46...21B,2010A&A...513A..79B} would hinder detection
of protoplanetary discs in the infrared, the only wavelength range available to detect signatures of individual circumstellar discs in the distant
Arches cluster. Whether grain growth in its disc population is expected to be as efficient as in the less massive star-forming regions in our solar
neighborhood depends on the environmental conditions. Recently, \citet{2011ApJ...731...95O,2011ApJ...731...96O} have shown that dust growth in discs
can be suppressed by significant ionization, a probable scenario given the strong X-ray irradiation of the numerous massive stars in the Arches
cluster \citep{2000ApJ...536..896C,2005MNRAS.361..679O}.

Second, if the radiation field of UV and X-ray photons emitted by massive stars is strong enough to heat the discs efficiently, photoevaporation will
drive additional serious disc-mass loss \citep{1998ApJ...499..758J,2004ApJ...611..360A,2007MNRAS.376.1350C,2009ApJ...705.1237G}. However,
\citet{2010ApJ...718..810S} have found a fraction of a few percent of A5V to B0V stars (corresponding to a mass range of $2-20\,\Msun$) in the Arches
cluster that show indisputable features of protoplanetary discs. Regarding its current age of 2.5\,Myr and enormous radiation field fed by more than
100~OB stars it seems that photoevaporation was not acting entirely destructive over this period.

So we expect that characteristic features of encounter-induced disc-mass loss -- such as the disc-poor tidal arms -- could be still apparent at the
present age of the Arches cluster. As the velocities of tidal tail stars do not differ by more than $2\sigma^O_{3D} \approx 20\,\kms$ from the large
space motion of the cluster of $\sim$200$\,\kms$ (see also Fig.~\ref{fig:tidal_diagnostics}), kinematically they can be well separated from the field
star population \citep[e.g.][]{2008ApJ...675.1278S}. This makes the Arches cluster a highly interesting observational target to study the combined and
distinct effect of photoevaporation and encounters on the evolution of protoplanetary discs in an extreme environment.

Note that observations of the stellar population and its circumstellar discs in this dense and distant cluster require instruments with very high
angular resolution $\theta \lesssim 0.1''$. Unlike successfully performed for the nearby, less dense young star clusters in the solar neighbourhood
\citep[e.g.][]{1999ARA&A..37..363F,2005ApJS..160..511K,2005ApJS..160..401P,2007A&A...464..211A,2009ApJ...696...47W}, members of the young Arches
cluster cannot be separated from the older field stars using the Chandra X-ray Observatory because of its limited resolution $\theta \gtrsim
0.5''$. Thus at present the only viable strategy to discern cluster members from field stars is kinematic selection based on long-term proper motion
studies.

A natural extension of such a study would include the $\sim$4\,Myr old, more extended Quintuplet cluster, that has been considered to be the Arches
cluster's ``older borther'' \citep{1999ApJ...525..750F}. Indeed, core collapse of our cluster model at 2.5\,Myr induces significant expansion (see
Fig.~\ref{fig:dynamics__isolated_vs_orbit}), naturally explaining the two starburst clusters as different evolutionary stages emerging from the same
initial conditions. Hence the Quintuplet cluster represents the ideal observational target to verify the future evolution of the Arches cluster as
predicted by our simulation. The characteristic concentration of disc-less stars at the edges of the extended tidal arms provides a critical test.

We have thus initiated a multi-epoch $K$-band campaign and $L$-band photometry of the Arches and the Quintuplet cluster to obtain a kinematically
selected membership sample required to observationally detect the characteristic disc-poor tidal features.

However, we note that near-infrared surveys are not ideal to trace the predicted spatial gradients in disk fractions. This is because this wavelength
regime is dominated by emission from the inner disc regions (at $\lesssim$1\,AU from the central star) while the encounter-induced disc-mass loss
studied here mostly removes material from the outer disc parts (at $\sim$100\,AU). Hence, ideally, observations would be carried out at millimeter
wavelengths. The instrument of choice for this purpose is ALMA as it provides the high spatial resolution ($\theta \approx 0.1''$) required to resolve
individual sources in the starburst clusters at the Galactic Centre.

%

\begin{acknowledgements}
  C.O. is grateful for extensive discussions with Andrea Stolte on this work. We thank the referee and the scientific editor Eric Feigelson for
  valuable comments and suggestions. C.O. appreciates funding by the German Research Foundation (DFG) grant OL 350/1-1, support by NAOC CAS through
  the Silk Road Project, and by Global Networks and Mobility Program of the University of Heidelberg (ZUK 49/1 TP14.8 Spurzem).
\end{acknowledgements}

%

\bibliographystyle{apj}
\bibliography{references}

%

\appendix

\section{Encounter Tracking}

\label{app:encounter_tracking}


\subsection{Numerical method}

\label{sec:numerics:modifications:encounter_tracking:method}

The essential task of the encounter tracking software is to determine the strongest perturber of each star at a given time step, i.e. the body with
the maximum gravitational force. Numerically this algorithm is solved by an effective oct-tree search \citep{2004physics...8067K}. This scheme,
repeated for each temporal data snapshot, provides the interaction period and the masses, positions and velocities of the perturber and the perturbed
star at their pericentre passage. While the interaction period can be easily obtained by recording the time step of the beginning and the end of the
interaction with the same perturber, the determination of the orbital parameters at periastron is more challenging.

\begin{figure}
  \begin{center}
    \includegraphics[width=0.5\linewidth]{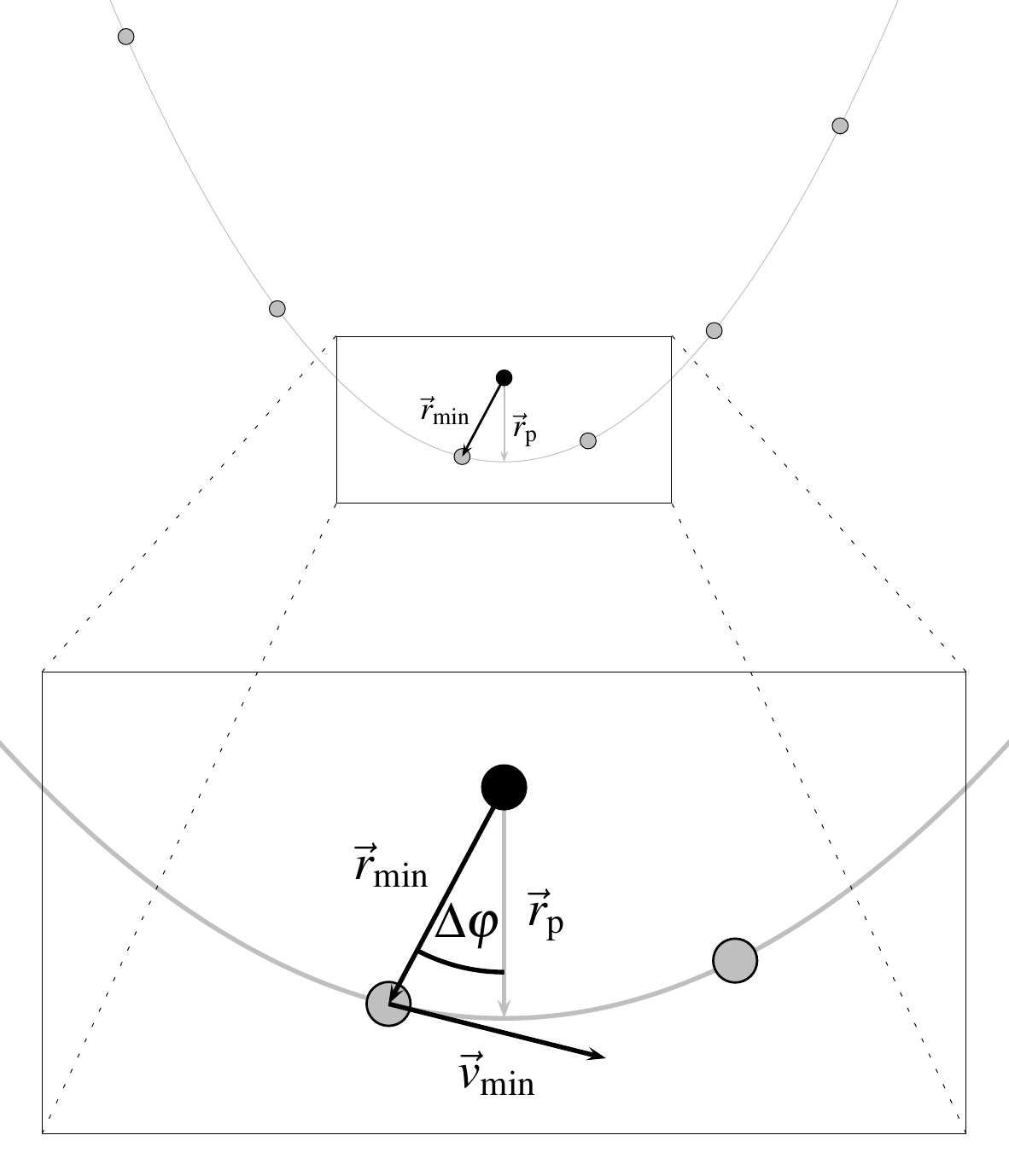}
    \caption{Illustration of the geometrical parameters of the periastron interpolation scheme. The black circle marks the position of the perturbed
      star, the grey circles mark the positions of the perturber at sequential time steps. The lower part represents an enlarged view and contains all
      quantities that are used for the interpolation scheme (see text).}
    \label{fig:numerics:periastron_interpolation}
  \end{center}
\end{figure}

However, for the calculation of the disc-mass loss of a star with mass~$m_1$ due to the fly-by of a perturber with mass~$m_2$, the determination of
their minimum distance -- the separation at periastron~$r_\mathrm{p}$ -- is crucial. Due to the discreteness of time in numerical simulations the
closest approach of two particles~$r_\mathrm{min}$ at a time step~$t_\mathrm{min}$ is usually not the minimum distance of their orbits~$r_\mathrm{p}$
as shown in Fig.~\ref{fig:numerics:periastron_interpolation}. As long as the interaction is weak and orbital periods are long the difference between
the numerical minimum separation and the theoretical distance at periastron is negligible. However, in case of strong interactions differences can be
quite large -- large enough that the estimated normalized disc-mass loss differs by more than a tenth or the orbit may be even classified as
non-regular (with respect to the criteria discussed below) and rejected. The solution is to interpolate the periastron distance~$r_\mathrm{p}$ at
time~$t_\mathrm{p}$ from the interaction parameters at some other time step as described in Sec.~\ref{sec:numerics:periastron_interpolation}.

Once the full interaction parameters are determined the orbit is checked for applicability of Eq.~\eqref{eqn:ImprovedFit} to calculate the disc-mass
loss. This involves two criteria: the orbit (i)~must be nearly Keplerian, and (ii)~must represent a significant part of the full Keplerian orbit. The
first criterion is satisfied if the eccentricities at the beginning, pericentre, and the end of the interaction do not differ by more than 10\,\% and
if the orbit has positive curvature (i.e. it is concave with respect to the perturbed star). The second criterion requires the opening angle of the
orbit section to amount to at least 10\,\% of the maximum opening angle of the corresponding conic section. Orbits that fulfil these criteria, termed
``regular'' in the following, are stored in a list that forms the backbone for the subsequent determination of the disc-mass loss.

\subsection{Periastron interpolation}

\label{sec:numerics:periastron_interpolation}

The interpolation of the periastron~$\vec{r}_\mathrm{p}$ requires the relative position and velocity vectors of a perturber, $\vec{r}$ and $\vec{v}$,
at a time step~$t$. We use the time step of the closest approach~$t_\mathrm{min}$ and assume that the orbit is (nearly) Keplerian, i.e. there is no
significant perturbation by a third body. This assumption is of course not always valid in the dense cluster regions, where encounters within small
$N$-body systems can occur. However, since the derived disc-mass loss is based on a parameter study that was restricted to single perturbers,
contributions from higher-order perturbations can not be quantified. So the disc-mass loss can be underestimated in this case.

Defining the total mass $M = m_1 + m_2$ and reduced mass $\mu = m_1 m_2 / M$, we use the six constants of motion given by the specific angular
momentum $h = \vec{r} \times \vec{v}$ and the Laplace-Runge-Lenz vector
\begin{equation*}
  \vec{A} = \vec{v} \times \vec{h} - \frac{GM}{r} \vec{r} \,,
\end{equation*}
where $G$ is the gravitational constant, to estimate the periastron distance~$r_\mathrm{p}$ and eccentricity~$\varepsilon$ of the orbit from
$\vec{r}_\mathrm{min}$ and $\vec{v}_\mathrm{min}$,
\begin{equation*}
  \begin{aligned}
    r_\mathrm{p} &= \frac{ |\vec{h}|^2 }{ GM\mu } \,,\\
    \varepsilon  &= \frac{ |\vec{A}| } { GM } \,.
  \end{aligned}
\end{equation*}
The interpolation of the time of periastron passage
\begin{equation*}
  t_\mathrm{p} = t_\mathrm{min} + \Delta{t} \,,\\
\end{equation*}
where $\Delta{t}$ is the time period it takes to move from $\vec{r}_\mathrm{min}$ to $\vec{r}_{\mathrm{p}}$ over the angle $\Delta\varphi =
\angle(\vec{r}_\mathrm{min},\vec{r}_\mathrm{p})$, is solved in polar coordinates $\{r,\varphi\}$. Again, we make use of the Laplace-Runge-Lenz vector
to determine~$\Delta\varphi$ via
\begin{equation*}
  \vec{A} \vec{r} = |\vec{A}| |\vec{r}| \cos{\varphi} \,.
\end{equation*}
From conservation of angular momentum,
\begin{equation*}
  h = r^2 \dot{\varphi} \,,
\end{equation*}
and the polar equation of a conic section,
\begin{equation*}
  r = \frac{(1 + \varepsilon) r_{\mathrm{p}}}{1 + \varepsilon \cos{\varphi}} \,,
\end{equation*}
one finds via separation of variables:
\begin{align*}
  \dot{\varphi} = & \frac{l}{r_p^2(1+\varepsilon)^2}(1+\varepsilon\cos{\varphi})^2 \\
  \quad \Longrightarrow \quad \text{d}t = & \frac{r_p^2(1+\varepsilon)^2}{l}\frac{\text{d}\varphi}{(1+\varepsilon\cos{\varphi})^2} \\
  \quad \Longrightarrow \quad \Delta{t} = & \frac{r_p^2(1+\varepsilon)^2}{l} \int_0^{\Delta{\varphi}} \frac{1}{(1+\varepsilon\cos{\varphi})^2}\text{d}\varphi \;.
\end{align*}
The result of the integration depends on the eccentricity $\varepsilon$,
\begin{equation*}
    \Delta{t} = \frac{r_p^2(1+\varepsilon)^2}{l}
    \begin{cases}
      \Biggl[ \frac{\varepsilon\sin{\Delta{\varphi}}}{(\varepsilon^2-1)(1+\varepsilon\cos{\Delta{\varphi}})} - \frac{1}{\varepsilon^2-1} \frac{2}{\sqrt{1-\varepsilon^2}}
      \tan^{-1}{\frac{(1-\varepsilon)\tan{\frac{\Delta{\varphi}}{2}}}{\sqrt{1-\varepsilon^2}}} \Biggr] & \text{if} \; \varepsilon<1 \,, \\
      \Biggl[ \frac{1}{2}\tan{\frac{\Delta{\varphi}}{2}} + \frac{1}{6}\tan^3{\frac{\Delta{\varphi}}{2}} \Biggr] & \text{if} \; \varepsilon=1 \,, \\
      \Biggl[ \frac{\varepsilon\sin{\Delta{\varphi}}}{(\varepsilon^2-1)(1+\varepsilon\cos{\Delta{\varphi}})} - \frac{1}{\varepsilon^2-1} \frac{1}{\sqrt{\varepsilon^2-1}}
      \ln{\frac{(\varepsilon-1)\tan{\frac{\Delta{\varphi}}{2}}+\sqrt{\varepsilon^2-1}}{(\varepsilon-1)\tan{\frac{\Delta{\varphi}}{2}}-\sqrt{\varepsilon^2-1}}} \Biggr] & \text{if}
      \; \varepsilon>1 \,.
    \end{cases}
\end{equation*}
Finally we need to determine whether $\vec{r}_\mathrm{min}$ was passed before or after $\vec{r}_{\mathrm{p}}$ in time, i.e. the sign of $\Delta{t}$,
which is given by the following relation:
\begin{equation*}
  \vec{r}_\mathrm{min} \vec{v}_\mathrm{min} \left\{
  \begin{aligned}
    < 0 \quad &\Rightarrow \quad \Delta\varphi \in (0,\pi)    &\Rightarrow \quad \Delta{t} > 0 \,,\\
    > 0 \quad &\Rightarrow \quad \Delta\varphi \in (\pi,2\pi) &\Rightarrow \quad \Delta{t} < 0 \,.
  \end{aligned}
  \right.
\end{equation*}
%

%

%

\end{document}